\RequirePackage{rotating} 
\documentclass[fleqn,usenatbib]{mnras}

\usepackage{newtxtext,newtxmath}

\usepackage[T1]{fontenc}

\DeclareRobustCommand{\VAN}[3]{#2}
\let\VANthebibliography\thebibliography
\def\thebibliography{\DeclareRobustCommand{\VAN}[3]{##3}\VANthebibliography}

\usepackage{letltxmacro}
\LetLtxMacro\oldttfamily\ttfamily
\DeclareRobustCommand{\ttfamily}{\oldttfamily\csname ttsize\endcsname}
\newcommand{\setttsize}[1]{\def\ttsize{#1}}%

\usepackage{bigints}
\usepackage{caption}
\usepackage{graphicx}
\usepackage{newtxtext,newtxmath}
\usepackage{mathtools}
\usepackage{amsmath}
\usepackage{dblfloatfix}

\makeatletter
\newcommand\notsotiny{\@setfontsize\notsotiny\@vipt\@viipt}
\makeatother

\title[M30 (NGC 7099) in the \textit{FUV}]{Far-ultraviolet investigation into the galactic globular cluster M30 (NGC 7099): I. Photometry and radial distributions}

\author[S. Mansfield et al.]{
Santana Mansfield$^{1}$\thanks{Email: smansfield@astro.uni-bonn.de},
Andrea Dieball$^{1}$,
Pavel Kroupa$^{1,2}$,
Christian Knigge$^{3}$,
David R. Zurek$^{4}$,
\newauthor
\ Michael Shara$^{4}$,
and Knox S. Long$^{5}$
\\ \\
$^{1}$Helmholtz-Institut f\"ur Strahlen- und Kernphysik (HISKP), Universit\"at Bonn, Nu\ss allee 14-16, 53115 Bonn, Germany\\
$^{2}$Astronomical Institute, Faculty of Mathematics and Physics, Charles University in Prague, V  Hole\v{s}ovi\v{c}k\'ach 2, 18000 Praha, Czech Republic\\
$^{3}$School of Physics and Astronomy, University of Southampton, Highfield, Southampton SO17 1BJ, UK\\
$^{4}$Department of Astrophysics, American Museum of Natural History, New York, NY 10024, USA\\
$^{5}$Space Telescope Science Institute, 3700 San Martin Drive, Baltimore, MD 21218, USA\\
}

\date{Accepted 2022 January 21; in original form 2021 October 26}

\pubyear{2022}

\begin{document}
\label{firstpage}
\pagerange{\pageref{firstpage}--\pageref{lastpage}}
\maketitle

\begin{abstract}
We present a far-ultraviolet (\textit{FUV}) study of the globular cluster M30 (NGC 7099). The images were obtained using the Advanced Camera for Surveys (ACS/SBC, F150LP, \textit{FUV}) and the Wide Field Planetary Camera 2 (WFPC2, F300W, \textit{UV}) which were both on board the \textit{Hubble Space Telescope (HST)}. The $FUV-UV$ colour-magnitude diagram (CMD) shows a main sequence (MS) turnoff at \textit{FUV} $\approx 22$ mag and $FUV-UV$ $\approx 3$ mag. The MS extends 4 mag below the turnoff, and a prominent horizontal branch (HB) and blue straggler (BS) sequence can be seen. A total of 1218 MS stars, 185 red giant branch stars, 47 BS stars and 41 HB stars are identified, along with 78 sources blueward of the MS which consist of white dwarfs (WDs) and objects in the gap between the WDs and the MS that include potential cataclysmic variable (CV) candidates. The radial distribution of the BS population is concentrated towards the cluster centre, indicating that mass segregation has occurred. The blue and red sub-populations of the double BS sequence appear mixed in the ultraviolet CMD, and no significant central concentration of CV candidates is seen in this cluster. 

\end{abstract}

\begin{keywords}
 ultraviolet: stars -- globular clusters: individual: M30 NGC 7099 
-- Hertzsprung-Russell and colour-magnitude diagrams --
                blue stragglers --
                techniques: image processing -- 
                techniques: photometric  
\end{keywords}


\setttsize{\small}

\section{Introduction}

\vspace{5pt}
Globular clusters (GCs) in our Milky Way are tightly bound groups of very old ($>\!\! 10$ Gyr) and metal-poor stars. They formed in the early Galaxy and can contain hundreds of thousands of stars, with extremely high stellar densities in the cores. At optical wavelengths, the luminosity of a GC is dominated by the large number of main sequence (MS) stars and evolved stars including red giant branch (RGB) stars and horizontal branch (HB) stars, as well as blue stragglers (BSs). Stellar exotica such as white dwarfs (WDs), binaries such as cataclysmic variables (CVs, consisting of a WD accreting mass from a companion) and low-mass X-ray binaries (LMXBs, containing a neutron star or a black hole accreting material from a low-mass companion) are optically faint and are therefore not easily detected in the cores of GCs in the visual wavebands. The spectral energy distribution of these hot exotic objects usually peaks at far-ultraviolet (\textit{FUV}) wavelengths, where MS stars and RGBs are faint. As a result, the core of a GC appears less crowded in the \textit{FUV} and the exotic sources can be more easily detected and examined. Several cores of Galactic GCs have been studied in the far-ultraviolet, including M2\ \citep{m2},\ M15\ \citep{m15},\ M80\ \citep{m80_2, m80}, NGC 1851 (\citealt{ngc1851, zurek2009, maccarone, zurek, Subramaniam2017}), NGC 2808 \citep{brown,ngc2808}, NGC 5466 \citep{Sahu}, NGC 6397 \citep{ngc6397}, NGC 6752 \citep{ngc6752}, 47 Tuc \citep{47tuc}. Also M3, M13 and M79 \citep{dalessandro2013}, and the outskirts of GCs have been observed with the \textit{Galaxy Evolution Explorer (GALEX)} survey of GCs \citep{Dalessandro2012, Schiavon2012}, and the \textit{Swift Gamma-Ray Burst Mission} UVOT survey \citep{Siegel_2014,Siegel_2015}.

We are able to probe deeper into GCs by detecting stellar populations that shine brightly in specific wavelengths, particularly those populations which formed by dynamical interactions. For example, blue stragglers (BS) appear to be on the MS above the turnoff point, although the cluster's age indicates that they should have evolved away from the MS to become red giants. These sources likely have mass added by either mass transfer from a companion \citep{mccrea}, or resulting from a direct collision \citep{hills}, which adds hydrogen fuel to the core allowing these stars to remain on the MS for longer than they would have otherwise. The high density of stars in the cluster core allows frequent interactions resulting in various exotic sources such as BS stars, CVs, X-ray binaries and millisecond pulsars (MSPs, \citealt{Shara_2006, Hurley_2007, ivanova, hong, kremer2020}). Detailed stellar-dynamical analyses can be found in \citet{leigh2007,leigh2011,leigh2013,leigh2015,chatterjee2010,chatterjee2013,hypki,wang2016,belloni2017,belloni2018a,belloni2018b,Belloni2019} and \citet{rui2021}. Binary systems are important for the dynamical evolution of the cluster \citep{heggie,hut}, as the binding energy in the binary can be transferred to other stars in dense stellar systems, stabilising the cluster against deep core collapse \citep{hills1975,hurley_shara2012,breen,rodriguez}, such that the initial binary population constitutes an essential aspect \citep{belloni2017}. Thus the detection of binary systems is important for our understanding of the dynamical state of the cluster.

M30 (NGC 7099) is a metal-poor globular cluster with a metallicity of [Fe/H] = --2.12 \citep{harris}. It is located 8.3\:$\pm$\:0.2\:kpc away from us in the Galactic halo and has an estimated age of 13.0\:$\pm$\:0.1\:Gyr \citep{kains}. M30 has a retrograde orbit with an average orbital eccentricity of $\langle e\rangle$ = 0.316, average perigalactic and apogalactic distances of $\langle r_{min}\rangle$ = 3.94 kpc and $\langle r_{max}\rangle$\:=\:7.58\:kpc respectively, and an average maximum distance from the Galactic plane of $\langle|z|_{max}\rangle$ = 4.95 kpc (\citealt{allen2006}, based on a nonaxisymmetric (barred) Galactic potential). M30 has a cluster mass of\:$\approx$\:1.6\:$\times$\:10$^5$\:M$_{\odot}$ and is core-collapsed: it has dynamically evolved so that the most massive stars have fallen inward and are concentrated towards the cluster core within a radius of only 1.9$''$\:(0.08\:pc, \citealt{sosin}). M30 has a very high central density ($\approx$\:10$^6$\:M$_{\odot}$\:pc$^{-3}$, \citealt{lugger}), indicating a high rate of stellar interactions and activity in the core region. This has resulted in a large bluer inward colour gradient, where the \textit{B$-$V} colour profile has a strong inclination towards bluer magnitudes in the centre, resulting from the infall of BS stars and a depletion of RGs from the core \citep{howell}. Optical studies on M30 have yet to find a significant CV or WD population.

Based on the optical \textit{Hubble Space Telescope (HST)} Wide Field Planetary Camera 2 (WFPC2) data, a significant BS population is observed in M30 \citep{guhathakurta}, and \citet{ferraro} suggest that these stars can be divided into two distinct blue straggler sequences in the \textit{V\,$-$\,I} colour-magnitude diagram (CMD). The blue BS sequence aligns with the zero-age main sequence (ZAMS) and the red sequence lies at a brighter magnitude of $\Delta V$\:$\approx$\:0.75\:mag. Isochrones from stellar evolution models representing collisions are found to lie along the blue BS sequence \citep{ferraro,sills}, and the red BSs populate a region that can be reproduced with models which undergo mass transfer \citep{xin}. \citet{zwart} also
produced stellar simulations and show that red BSs form by mass transfer at a constant rate over the cluster lifetime, whereas the blue BSs may have formed by mergers during a short burst of activity when the core of the cluster collapsed roughly 3.2 Gyr ago. The burst of activity would have resulted from the increased likelihood for interactions between stars due to the collapsing cluster core. It remains unknown however why the two sequences are parallel and why the red sequence is\:$\approx$\:0.75\:mag brighter than the blue one. Double BS sequences have also been detected in NGC 362 \citep{Dalessandro_2013} and NGC 1261\citep{Simunovic_2014}.

This first work in our study of M30 aims at analysing \textit{HST FUV} and $UV$ data using photometry and providing an ultraviolet CMD and an analysis of the radial distributions of the different stellar populations. The observations and data reduction are described for each filter in Sect. \ref{observation} and the \textit{FUV} $-$ \textit{UV} CMD is presented in Sect. \ref{cmd}. The optical counterpart matching is given in Sect. \ref{opticalsec}, and radial distribution analysis in Sect. \ref{radialsec}, followed by a summary in Sect. \ref{summ}.
A following publication in this study will include potential counterparts to known X-ray sources and an investigation into the variable sources in M30.

\section{Observations and image processing}
\label{observation}

\begin{table*}
    \centering
 \caption{Log of the observation dates, the camera and filters used, and the total exposure times. \label{obslog}}
\begin{footnotesize}
\begin{tabular}{ccccc}
\hline
Camera&Filter&Data Set&Start Date&Exposure Time (s)\\
\hline\hline
    ACS/SBC&F150LP&J9HC02011&29 May 2007 16:45:50&5040\\
    ACS/SBC&F150LP&J9HC04011&03 June 2007 11:51:14&5040\\
    ACS/SBC&F150LP&J9HC04021&03 June 2007 15:01:42&2520\\
    ACS/SBC&F150LP&J9HC06011&09 June 2007 02:10:34&7560\\
	WFPC2&F300W&U9HC0301M-U9HC0308M&29 May 2007 19:58:16&4000\\
	WFPC2&F300W&U9HC0501M-U9HC0508M&03 June 2007 16:39:16&4000\\	 
	WFPC2&F300W&U9HC0701M-U9HC0708M&09 June 2007 07:00:16&4000\\

\hline
\end{tabular}
\end{footnotesize}
    \label{observations}
\end{table*}

\begin{figure}
    \centering
    \includegraphics[width=\columnwidth]{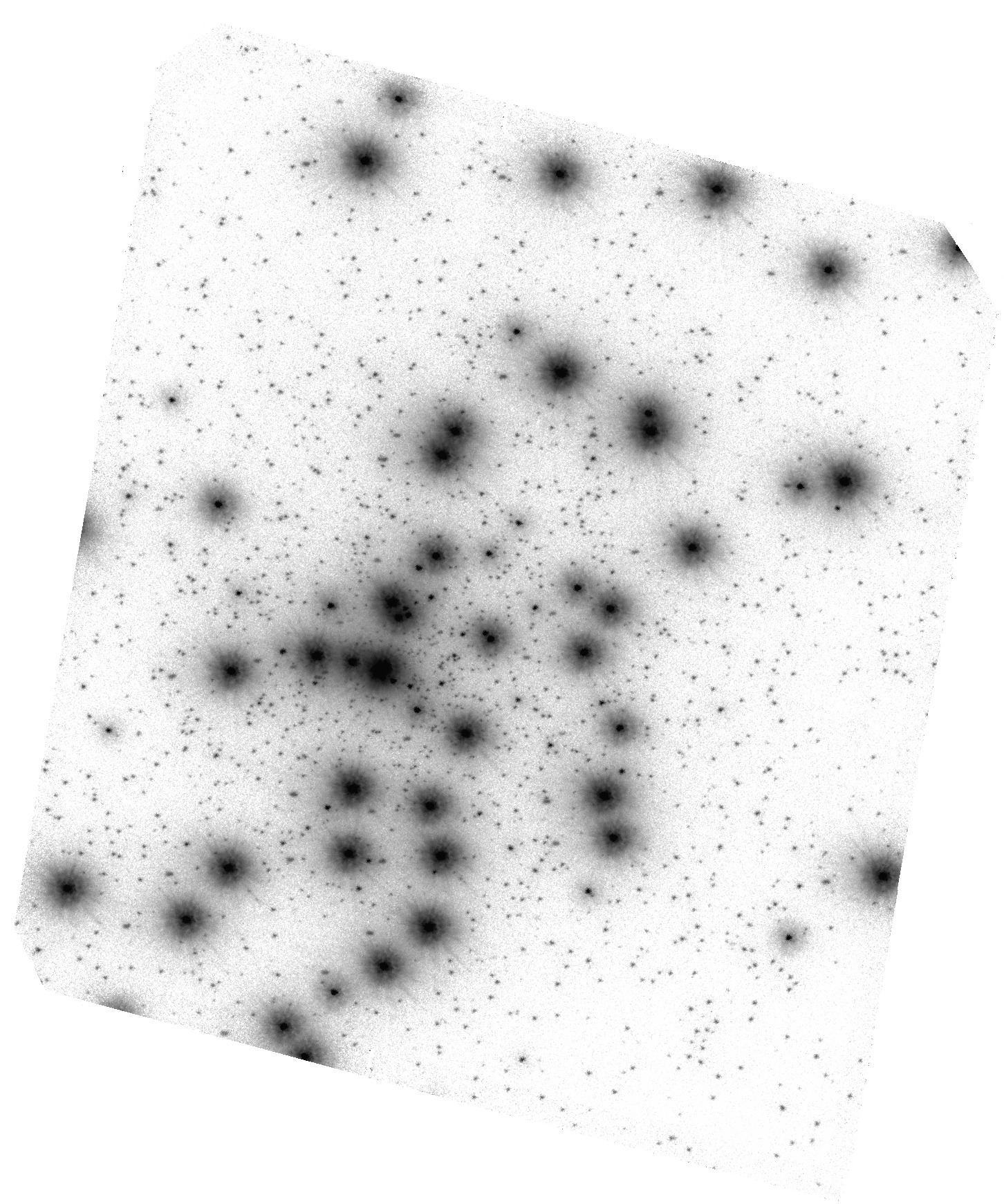}
    \caption{Master image of the \textit{FUV} ACS/F150LP exposures of the core of M30, which spans 34.6$''$ $\times$ 30.8$''$ (1.40 pc $\times$ 1.24 pc). 
    North is up and east is to the left.}
    \label{fuv}
\end{figure}

\vspace{5pt}
\subsection{The ACS \textit{FUV} data}

\vspace{5pt}
The far-ultraviolet (\textit{FUV}) data was obtained with the Advanced Camera for Surveys (ACS) on board the \textit{HST}, using the Solar Blind Channel (SBC) and F150LP filter (program GO-10561; PI: Dieball), in 15 orbits distributed over three visits,
from the 29th May to the 9th June 2007. The SBC has a field of view of 34.6$''$ $\times$  30.8$''$ and a pixel scale of 0.034$''$ $\times$ 0.030$''$ pixel$^{-1}$. 
Sixty-four images were taken, each with an exposure time of 315 s, resulting in a total exposure time of 20160 s, and are described in Table \ref{obslog}. 
Thermal breathing during the 96 minute day-night cycle of the \textit{HST} for each visit, as well as guide star re-acquisition,
can create small shifts between the individual images taken at the same pointing position. As a first step, a master image is created  which realigns all individual images using the \texttt{TWEAKREG} task from the \texttt{DRIZZLEPAC} package running under \texttt{PYRAF} \citep{pyraf}. This task compares the world coordinate system (WCS) data in the header of each image to a reference image (the first in our list) and calculates the small residual shifts made by the pointing differences.
The threshold is adjusted to ensure the rms of the shifts is small and there is no correlation in the residual plots. 
The images are then combined into a geometrically-corrected master image using the \texttt{ASTRODRIZZLE} task from the \texttt{DRIZZLEPAC} package. This task incorporates the processes of performing sky subtraction, drizzling, creating a median image, cosmic ray removal and final combination into the master image. In this instance, cosmic ray removal is not needed as the SBC is not sensitive to cosmic rays. The \textit{FUV} master image is shown in Fig.~\ref{fuv}. Although this shows the extremely dense core region of the cluster, the \textit{FUV} image does not suffer from much crowding since the numerous MS stars are faint at these wavelengths.

\subsubsection{Object identification}

\vspace{5pt}
Most of the potential sources are automatically detected using the \texttt{DAOFIND} task in the \texttt{DAOPHOT} package \citep{stetson}, running under \texttt{PYRAF}. This task determines if there is a source in a given pixel by applying a Gaussian profile on the surrounding pixels. A detection is recorded if there is a good fit with a large positive central height of the Gaussian, whereas a negative or minimal height resembles the edge of a star or an empty region of sky \citep{stetson}. A FWHM\:of\:3 and zero readnoise is used for the ACS/SBC. The coordinates of found sources are over-plotted onto the master image to be checked by eye. The threshold for \texttt{DAOFIND} needs to be chosen such that it detects as many faint sources as possible without making too many false detections in the vicinity of the brightest stars. These false detections are then removed from the list by hand (along with some false detections at the edge of the image), and any of the faint sources missed by the \texttt{DAOFIND} task are added. The resulting number of objects found in the \textit{FUV} is 1934. 

\subsubsection{Stellar photometry}

\vspace{5pt}
Aperture photometry was performed using the task \texttt{DAOPHOT} \citep{stetson} running under \texttt{PYRAF}. An aperture of 4 pixels was used with a sky annulus of 8-12 pixels. A small aperture is needed for dense stellar systems, however in order to account for the limited percentage of flux contained in the small aperture, an aperture correction, ApCorr, is applied, which is the inverse of the encircled energy. The small sky annulus also contains light from the star itself,  and thus a sky correction, SkyCorr, is needed. Assuming that a larger sky annulus of 60-70 pixels contains only background flux, SkyCorr is determined by taking the magnitude ratio of the large sky annulus and the small sky annulus for several chosen isolated stars (see e.g. \citealt{m15}). The fluxes are then converted into STMAGs using the formula:

$$\textnormal{STMAG =  - 2.5 $\times$ log (flux $\times$ SkyCorr $\times$ ApCorr / ExpTime)}$$

\vspace*{-15pt}
$$\textnormal{+ ZPT},$$

\noindent where ZPT = --21.1 mag is the zero point for the STMAG system, and flux = counts $\times$ PHOTFLAM, where PHOTFLAM is the inverse sensitivity of the instrument used to convert count rates into fluxes, and can be found in the image header. 
For the ACS filter F150LP, this is:

\vspace*{-10pt}
\begin{equation*}
 \rm{PHOTFLAM}_{ACS,F150LP} =  3.246305 \times 10^{-17} \rm{erg}\ \rm{cm}^{-2}\ \textup{\AA}^{-1}\ \rm{count}^{-1}   
\end{equation*}

\vspace{5pt}
\noindent For an annulus of 4 pixels, the corrections are ApCorr\:=\:1.7636 \citep{acs} and SkyCorr = 1.011. 

\subsection{The WFPC2 F300W data}

\begin{figure}
    \centering
    \includegraphics[width=\columnwidth]{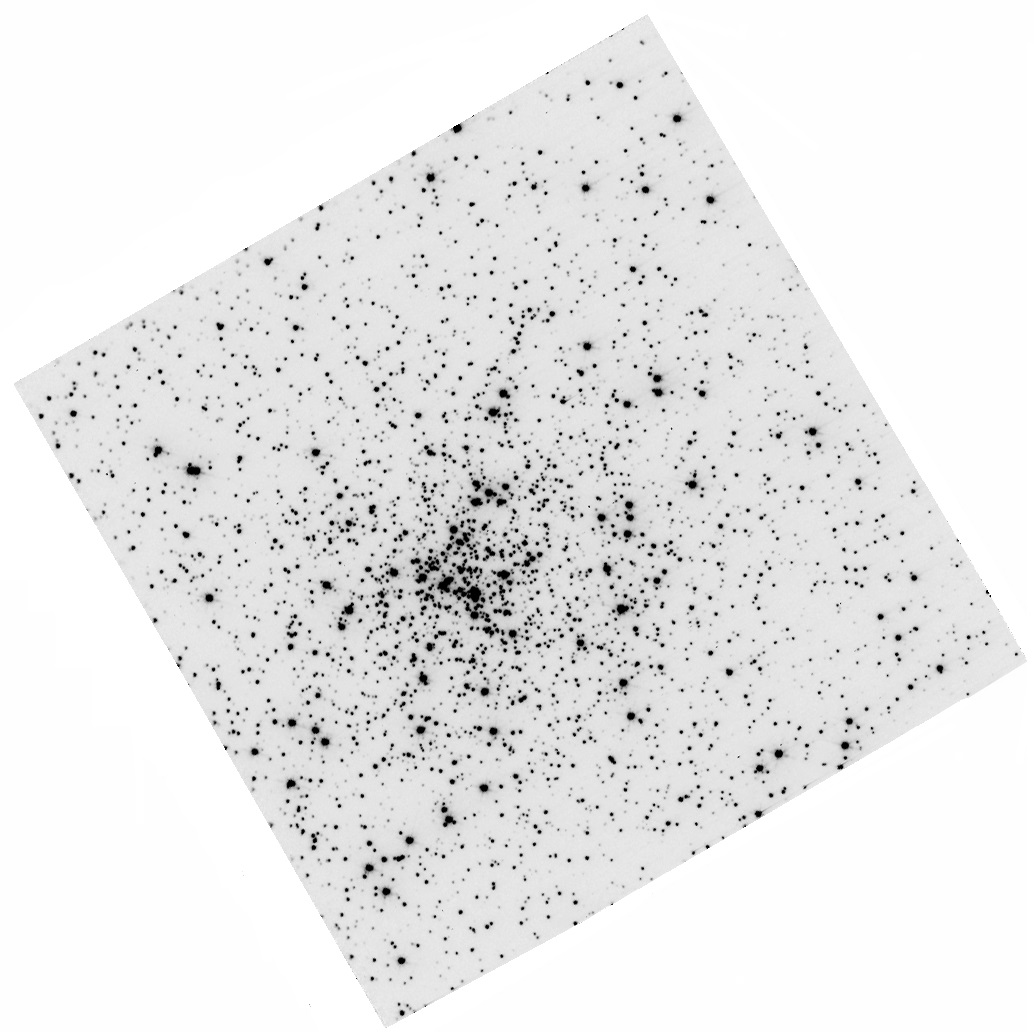}
    \caption{F300W master image of the WFPC2/PC exposures of M30, with a field of view of $34\arcsec \times 34\arcsec$ (1.37 pc $\times$ 1.37 pc).  North is up and east is to the left.}
    \label{UV}
\end{figure}

\vspace{5pt}
The \textit{UV} data was obtained using the WFPC2 which was on board the \textit{HST} (program GO-10561; PI: Dieball) with the F300W filter in 15 orbits distributed over three visits,
from the 29th May to the 9th June 2007. The WFPC2 consists of four cameras, the three Wide Field Cameras (WFCs) each with a field of view of $50'' \times 50''$, and the Planetary Camera (PC) with a field of view of $34'' \times 34''$, and a pixel scale of 0.046$''$ pixel$^{-1}$. The PC was centred on the cluster core. The exposures from these four chips are placed together into a mosaic image. Twenty-four mosaic images were taken with a total exposure time of 12000 s (Table \ref{observations}).

 \begin{figure*}
    \centering
    \includegraphics[scale=0.54]{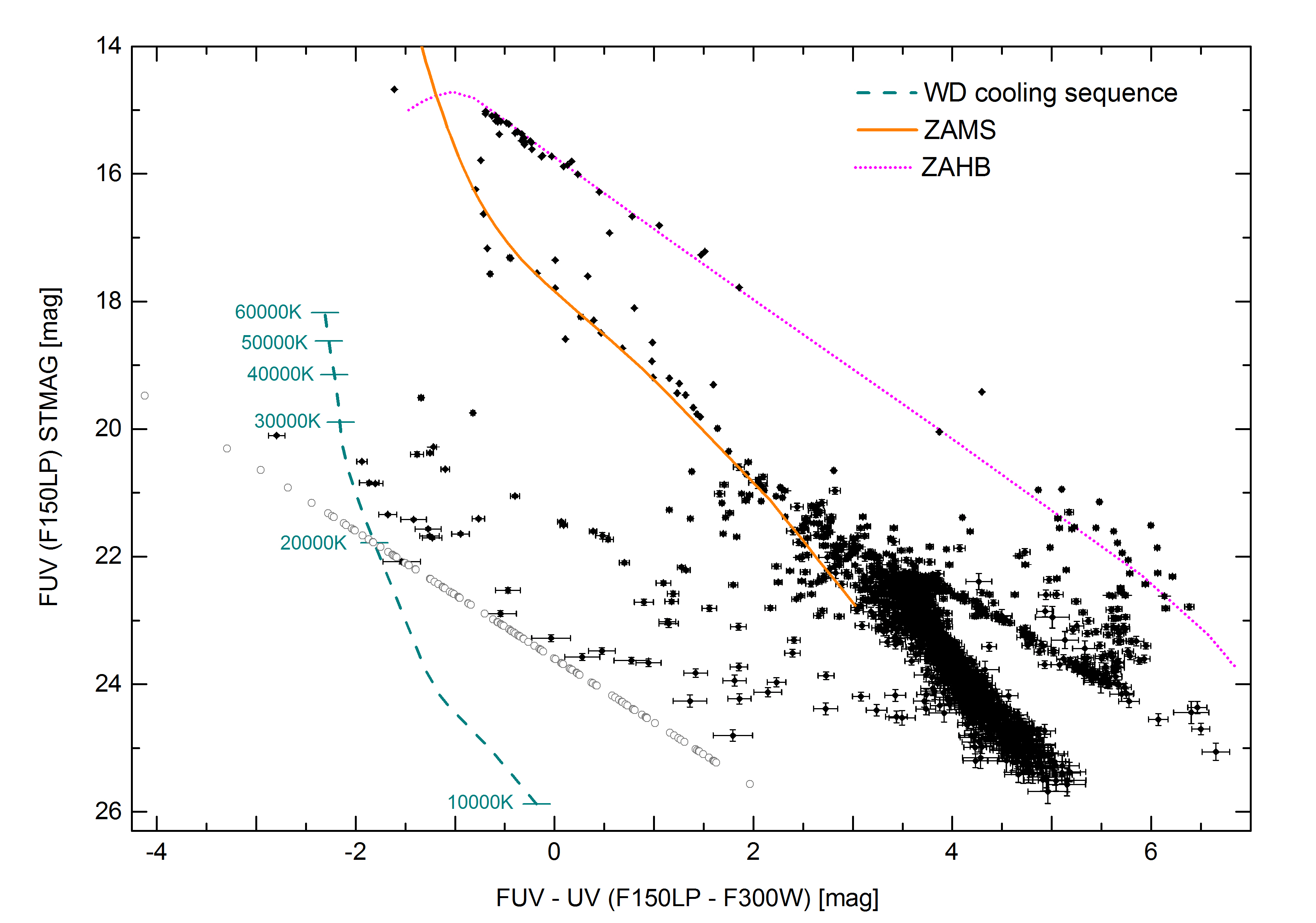}
    \caption{\textit{FUV} -- \textit{UV} CMD of M30. A theoretical ZAMS (solid orange line), ZAHB (dotted pink line), and a CO WD cooling sequence (dashed green line) have been added for orientation. The surface temperatures of WDs are also indicated. Additionally shown are sources measured in the \textit{FUV} but with no \textit{UV} counterpart (grey circles), using the detection limit of the \textit{UV} to estimate their position on the CMD, which also represents the location of the detection limit.}
    \label{isochrones}
\end{figure*}

\subsubsection{Object identification and stellar photometry}

\vspace{5pt}
A master image is created using the same method that was used for the \textit{FUV} data, with the \texttt{TWEAKREG} and \texttt{ASTRODRIZZLE} tasks from the \texttt{DRIZZLEPAC} package running under \texttt{PYRAF} \citep{pyraf}, and is shown in Fig.~\ref{UV}. Photometry was performed using the package \texttt{DOLPHOT} \citep{dolphot}, running under \texttt{IRAF} \citep{iraf1,iraf2}. This program carries out the image alignment with respect to a reference image (our master image), source detection and  photometry directly on the 24  flat-fielded exposures. \texttt{DOLPHOT} contains a specific module for the WFPC2 which performs PSF-fitting using a pre-calculated PSF model for the F300W filter\footnote{\url{http://americano.dolphinsim.com/dolphot}}, and the output magnitudes are already flux corrected. As the ACS/\textit{FUV} field of view is contained within the field of view of the Planetary Camera, only the data from this chip is needed here. For the photometry, the recommended parameters for the WFPC2 module are first used, then optimised so that as many faint sources were measured as possible\footnote{The optimised parameters are: imgRPSF = 10, RCentroid = 1}. The number of objects found in the PC/F300W data is 10451, with 8836 of these within the \textit{FUV} field of view.

\subsection{Catalogue matching}{\label{transformation}}

\vspace{5pt}
In order to find objects that appear in both the \textit{FUV} and \textit{UV} data, the coordinates of the sources found in the \textit{FUV} image are transferred into the \textit{UV} frame. In order to make the conversion, the $x$ and $y$ positions of 30 reference stars easily identified in both images are used as input for the task \texttt{GEOMAP} running under \texttt{PYRAF}. This computes the spatial transformation which is used by the \texttt{GEOXYTRAN} task to transform the coordinates of the \textit{FUV} objects into the \textit{UV} frame. This is checked by over-plotting the transformed \textit{FUV} coordinates onto the \textit{UV} image. The two catalogues are then compared for matching stars using the task \texttt{TMATCH} which correlates the two lists for sources matching in coordinates within a radius of a given number of pixels. The average matching radius for sources in both frames is 0.35 pixels, and this procedure results in 1569 matching objects. The first 30 entries of the final list of sources are given in Table \ref{fullcat}.

A number of chance matches are expected, and can be estimated using the numbers of objects in the two wavelengths and the matching radius \citep{knigge}. The estimated number of spurious matches is 74 pairs (4.6$\%$ of all matches).

\section{The \textit{FUV} -- \textit{UV} CMD} \label{cmd}

\vspace{5pt}
The \textit{FUV} -- \textit{UV} CMD is given in Fig. \ref{isochrones}. A theoretical zero-age main sequence (ZAMS), zero-age horizontal branch (ZAHB), and WD cooling sequence with marked surface temperatures are included for orientation purposes. These are calculated using the fitting formulae of \citet{tout}, the theoretical ZAHB models from \citet{dorman} and the \citet{wood} grid of WD cooling curves with a grid of synthetic WD spectra by \citet{gansicke}. Kurucz models of stellar atmospheres are used for interpolation and the resulting spectra are then folded with the appropriate filter and detector combinations using \texttt{PYSYNPHOT} running under \texttt{PYRAF}. The cluster parameters used are a distance of 8.3 kpc, a metallicity of [Fe/H]$=-2.12$ and a reddening of $E(B-V)=0.03$ mag.

\begin{figure}
  \centering
    \includegraphics[scale=0.13]{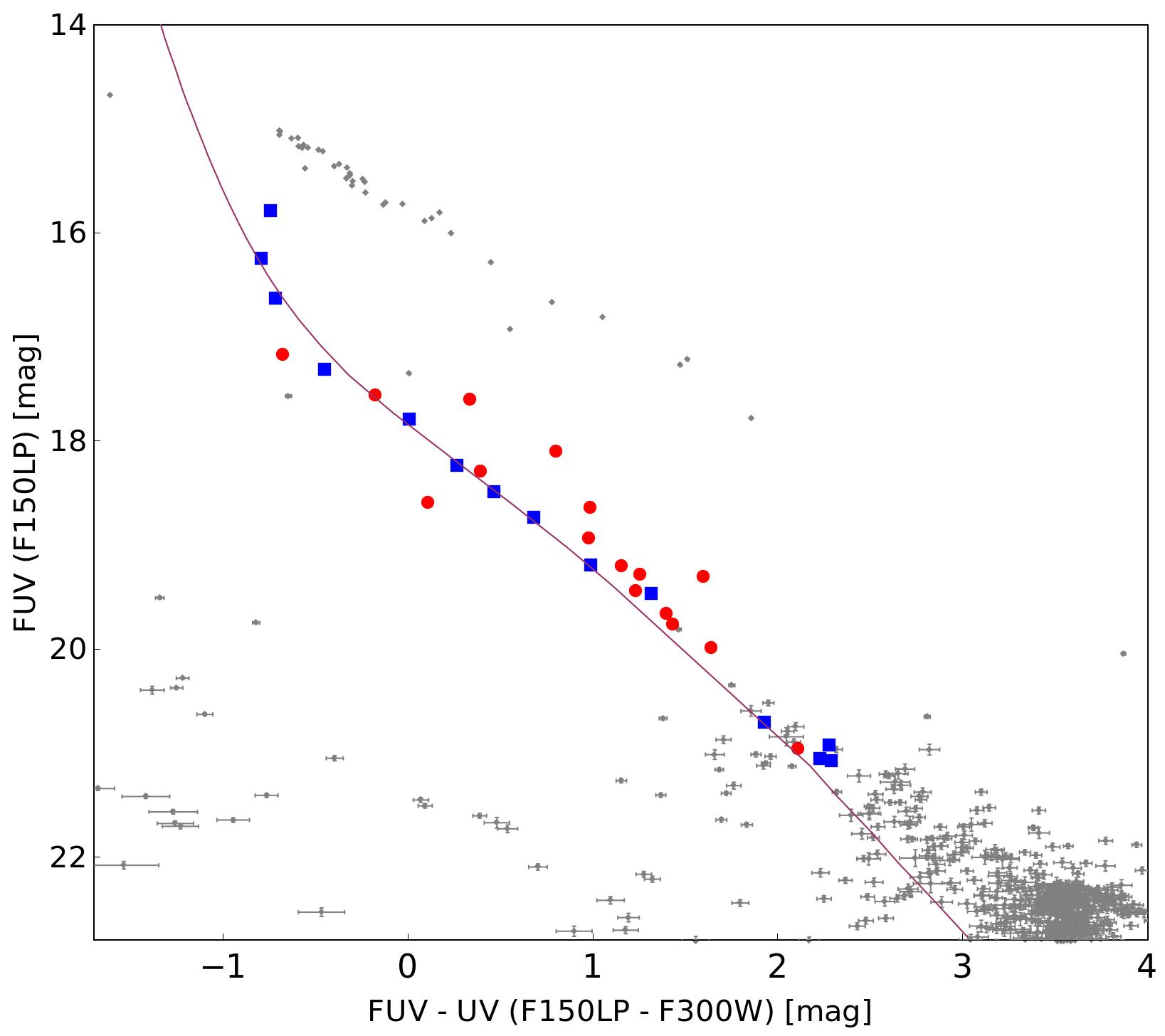}
    \includegraphics[scale=0.13]{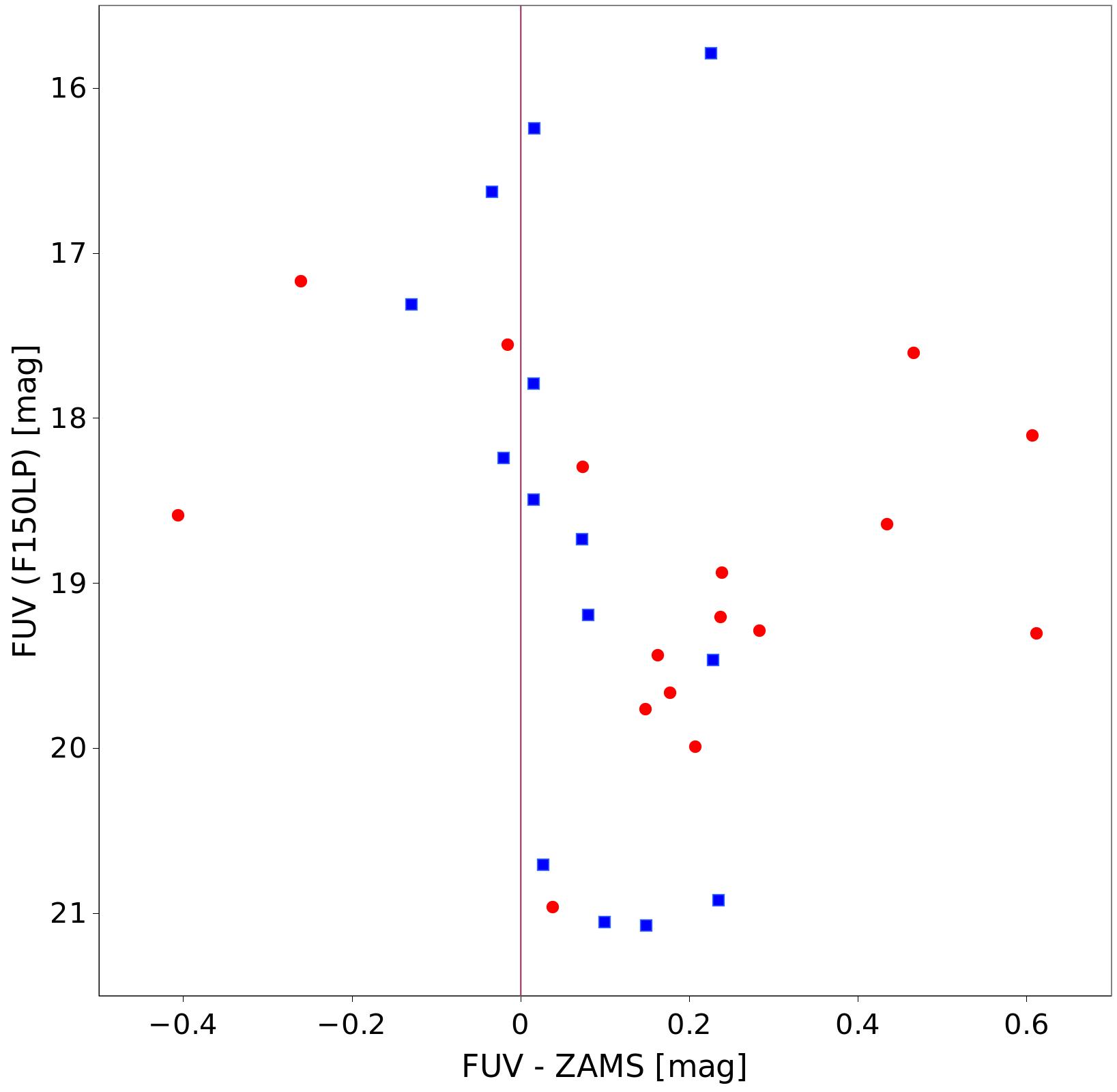}
    \caption{The BS sources that are found to be in a double BS sequence in the infrared by \citet{ferraro}, marked according to whether they belonged to the blue (squares) or red (circles) BS sequence, on the $FUV - UV$ CMD (top) and their position verticalized along with respect to the ZAMS (bottom)}. \label{ferraro}
\end{figure}

There is a well-defined main sequence, with turnoff at $FUV \approx 22\ \textnormal{mag}$ and \textit{FUV} -- \textit{UV} $\approx$ 3 mag. Many BS sources are seen along the ZAMS at brighter magnitudes than the turnoff. The RGB lies from the turnoff towards fainter and redder magnitudes. The HB reaches from the RGB up to \textit{FUV} $\approx$ 15 mag. The sources between the MS and WD cooling sequence are called gap objects and are likely to include CV candidates \citep{m15}. Also included are 126 sources present in both $FUV$ and $UV$ fields of view with $FUV$ measurements but no $UV$ counterparts. The detection limit of $UV = $ 23.6 mag was used to estimate the position of these sources in the CMD, and as such they lie on the line representing this limit. 

Figure \ref{ferraro} shows 16 of the BSs that are situated in the red BS sequence in \citet{ferraro}, and 13 from the blue BS sequence. As the separations in colour are small, they are also plotted in Fig. \ref{ferraro} in a verticalized CMD with respect to the theoretical ZAMS curve used in Fig. \ref{isochrones}. The highest uncertainty in the  \textit{FUV - UV} position for these sources was 0.024 mag, and 0.004 mag for the $FUV$. Of the two BS sources without counterparts from \citet{ferraro}, the source at $FUV - UV \approx 0$ mag (ID27) was just outside of their field of view, and the one at $FUV - UV \approx -\ 0.65$ mag (ID283) matched within 2 pixels to the position of a source that was between the double BS sequence and the RGB (\citealp{ferraro}, their source \#11002543), and as such was not included. The field of view in our study only included the cluster centre, thus further observations using a larger field of view of this cluster in \textit{FUV} wavelengths is needed to fully explore the BS population, however we find that there is no clear distinction between the two BS sequences in our dataset. The sources that were attributed to the distinct sequences in \citet{ferraro} appear to be well mixed in the ultraviolet, particularly as two of the red BS sources have a large blue excess compared to the other BSs (Fig. \ref{ferraro}). As the sources on the red BS sequence are thought to be the result of mass transfer \citep{xin}, the reason the two sub-populations appear mixed may be due to the sources on the red sequence being in an active state of mass transfer. \citet{xin} detail a binary model which begins mass transfer at 7.54 Gyr, becomes brighter than the MS turnoff at around 10 Gyr, remains in the BS region for another 3-4 Gyr with mass transfer ceasing at 12.78 Gyr. Systems with active mass transfer emit $FUV$ radiation due to the hot material in the accretion disk which would give these sources a blue excess in the $FUV - UV$ CMD. These active systems  would then be shifted bluewards towards (and even beyond) the blue BS sequence in the ultraviolet.

\captionsetup[table]{labelfont={bf},name={Table},labelsep=period}

\begin{sidewaystable*}

\caption{The first 30 entries from the catalogue of sources in the \textit{FUV} and \textit{UV} field of view. The ID number is given in the first column, the position of the source in Cols. 2 and 3, the radial distance is given in Col. 4, and the pixel position of the source on the \textit{FUV} image is in Cols. 5 and 6. The magnitudes are in Cols. 7 $-$ 9 and the resulting population type of each source (determined from the position in the \textit{FUV} -- \textit{UV} CMD) is given in Col. 10. Also included in Cols. 11$-$13 are the corresponding $B - V$ and $V$ magnitudes and ID number from the optical catalogue \citep{guhathakurta}. This table is published in its entirety in the supplementary material (online).}

\small{
\begin{tabular}{ccccrrccccrcr}
\hline 

ID & RA & Dec & $r^a$ & $x_{FUV}$ & $y_{FUV}$ & \textit{FUV} & \textit{UV} & \textit{FUV}$-$\textit{UV} & Type & B$-$V$^b$ & V$^b$ & ID$_{Optical}^b$\\
 & [h:m:s] & [$^{\circ}$:$'$:$''$] & [arcmin] & [pixels] & [pixels] & [mag] & [mag] & [mag] & & [mag] & [mag]\\ 
\hline\hline \\
1&21:40:20.94&-23:10:55.71&0.313&479.48&74.20&15.431\ $\pm$\ 0.001&15.742\ $\pm$\ 0.001&-0.311\ $\pm$\ 0.002&HB&-0.09&15.63&755\\
2&21:40:20.79&-23:10:48.11&0.314&794.64&81.69&22.464\ $\pm$\ 0.030&18.868\ $\pm$\ 0.004&3.596\ $\pm$\ 0.034&MS& 0.34&18.69&635\\    	
3&21:40:20.87&-23:10:48.48&0.296&767.87&119.88&22.654\ $\pm$\ 0.033&18.908\ $\pm$\ 0.004&3.746\ $\pm$\ 0.037&MS&0.37&18.74&699\\	4&21:40:20.89&-23:10:47.74&0.291&793.36&137.83&22.550\ $\pm$\ 0.039&18.654\ $\pm$\ 0.004&3.896\ $\pm$\ 0.036&MS&0.39&18.43&714\\	
5&21:40:20.81&-23:10:42.51&0.319&1005.74&157.13&22.583\ $\pm$\ 0.033&18.852\ $\pm$\ 0.004&3.731\ $\pm$\ 0.037&MS&0.36&18.67&665\\
6&21:40:20.92&-23:10:46.79&0.282&823.92&168.39&22.521\ $\pm$\ 0.031&18.835\ $\pm$\ 0.004&3.686\ $\pm$\ 0.035&MS&0.34&18.69&746\\
7&21:40:21.19&-23:10:57.86&0.283&358.25&182.20&18.495\ $\pm$\ 0.005&18.027\ $\pm$\ 0.003&0.468\ $\pm$\ 0.008&BS&&&\\
8&21:40:21.16&-23:10:56.00&0.270&434.47&187.06&21.511\ $\pm$\ 0.020&15.513\ $\pm$\ 0.002&5.998\ $\pm$\ 0.022&RG&1.12&12.69&1005\\
9&21:40:21.16&-23:10:54.62&0.256&485.79&207.67&22.478\ $\pm$\ 0.031&18.767\ $\pm$\ 0.004&3.711\ $\pm$\ 0.035&MS&0.44&18.57&1025\\	
10&21:40:20.90&-23:10:42.02&0.301&1010.36&210.87&22.364\ $\pm$\ 0.032&18.781\ $\pm$\ 0.004&3.583\ $\pm$\ 0.036&MS&0.29&18.57&730\\
11&21:40:21.22&-23:10:56.63&0.264&400.37&213.90&22.314\ $\pm$\ 0.029&16.099\ $\pm$\ 0.001&6.215\ $\pm$\ 0.030&RG&0.72&14.65&1107\\	
12&21:40:21.30&-23:10:58.23&0.266&325.55&240.04&22.390\ $\pm$\ 0.030&18.784\ $\pm$\ 0.004&3.606\ $\pm$\ 0.034&MS&&&\\
13&21:40:20.97&-23:10:40.17&0.298&1071.16&266.28&22.401\ $\pm$\ 0.033&18.685\ $\pm$\ 0.004&3.716\ $\pm$\ 0.037&MS&0.33&18.48&796\\
14&21:40:21.26&-23:10:53.33&0.226&520.49&272.55&21.896\ $\pm$\ 0.023&18.325\ $\pm$\ 0.003&3.571\ $\pm$\ 0.026&MS&0.36&18.16&1163\\
15&21:40:21.23&-23:10:51.75&0.223&586.04&273.16&22.376\ $\pm$\ 0.029&18.568\ $\pm$\ 0.004&3.808\ $\pm$\ 0.033&MS&0.34&18.29&1120\\	
16&21:40:21.19&-23:10:50.02&0.223&658.10&274.11&22.578\ $\pm$\ 0.032&18.921\ $\pm$\ 0.004&3.657\ $\pm$\ 0.036&MS&0.36&18.82&1070\\
17&21:40:21.03&-23:10:42.37&0.270&976.93&274.68&19.812\ $\pm$\ 0.009&18.346\ $\pm$\ 0.003&1.466\ $\pm$\ 0.012&BS&0.12&18.18&860\\
18&21:40:21.01&-23:10:41.41&0.280&1016.38&276.28&15.187\ $\pm$\ 0.001&15.758\ $\pm$\ 0.001&-0.571\ $\pm$\ 0.002&HB&-0.12&15.78&841\\
19&21:40:21.36&-23:10:54.51&0.216&459.70&312.16&22.301\ $\pm$\ 0.028&18.568\ $\pm$\ 0.004&3.733\ $\pm$\ 0.032&MS&0.57&18.27&1278\\
20&21:40:21.31&-23:10:52.23&0.207&554.79&311.87&22.431\ $\pm$\ 0.031&16.489\ $\pm$\ 0.001&5.942\ $\pm$\ 0.032&RG&0.69&15.14&1222\\
21&21:40:21.23&-23:10:37.61&0.267&1127.68&436.35&22.270\ $\pm$\ 0.028&23.859\ $\pm$\ 0.276&-1.589\ $\pm$\ 0.304&Gap&0.30&18.70&1221\\		
22&21:40:21.19&-23:10:45.25&0.222&841.07&327.62&22.445\ $\pm$\ 0.030&18.627\ $\pm$\ 0.004&3.818\ $\pm$\ 0.034&MS&0.37&18.33&1075\\
23&21:40:21.03&-23:10:37.57&0.307&1161.20&328.22&22.492\ $\pm$\ 0.032&18.604\ $\pm$\ 0.004&3.888\ $\pm$\ 0.036&MS&0.43&18.32&865\\
24&21:40:21.25&-23:10:47.84&0.205&732.63&329.89&22.497\ $\pm$\ 0.031&18.814\ $\pm$\ 0.004&3.683\ $\pm$\ 0.035&MS&0.32&18.70&1158\\	
25&21:40:21.47&-23:10:57.80&0.235&316.94&331.07&22.390\ $\pm$\ 0.030&18.640\ $\pm$\ 0.004&3.750\ $\pm$\ 0.034&MS&0.40&18.49&1459\\
26&21:40:21.16&-23:10:43.49&0.235&913.04&333.06&22.618\ $\pm$\ 0.034&18.496\ $\pm$\ 0.003&4.122\ $\pm$\ 0.037&RG&0.37&18.26&1036\\
27&21:40:21.13&-23:10:41.56&0.254&992.76&335.06&17.353\ $\pm$\ 0.003&17.345\ $\pm$\ 0.002&0.008\ $\pm$\ 0.005&BS&-0.01&17.37&987\\
28&21:40:21.12&-23:10:40.56&0.262&1031.70&345.38&22.304\ $\pm$\ 0.034&18.767\ $\pm$\ 0.004&3.537\ $\pm$\ 0.038&MS&0.32&18.70&983\\
29&21:40:21.48&-23:10:56.31&0.213&371.12&357.70&22.529\ $\pm$\ 0.032&18.961\ $\pm$\ 0.004&3.568\ $\pm$\ 0.036&MS&0.35&18.77&1482\\
30&21:40:21.35&-23:10:49.55&0.185&651.59&362.87&21.075\ $\pm$\ 0.016&18.783\ $\pm$\ 0.004&2.292\ $\pm$\ 0.020&BS&0.18&18.69&1269\\
:& & & & & & & & & & & & \\

\hline

\end{tabular}}\label{fullcat}

\vspace{5pt}

(a) From the cluster centre at $x$ = $443.9$, $y$ = $362.8$ in the \textit{UV} image, corresponding to RA = $21^h40^m22\fs13,$ Dec = $- 23^{\circ}10'47\farcs40,$ determined as the centre of gravity by \citet{ferraro}.\\
(b) \citet{guhathakurta}.\\

\end{sidewaystable*}

There are around 78 WD/Gap objects blueward of the main sequence. We can make an estimate of the expected number of WDs in M30 by assuming that the number of stars in both of the HB and WD phases is proportional to the lifetime of these phases \citep{Richer1997}:

\begin{equation}
    \frac{N_{WD}}{N_{HB}} \approx \frac{\tau_{WD}}{\tau_{HB}}
\end{equation}

\noindent If we assume that $\tau_{HB} \approx 10^8$ yr \citep{dorman1992a}, the temperature of the WD cooling curve at the detection limit is 20,000 K $\approx 5 \times 10^7 \textnormal{yr}$ \citep{althaus}, and using the number of HB sources found in our sample $N_{HB} = 41$, we find that the estimated expected number of WDs is $N_{WD} \approx 20$. This rough estimate is in agreement with the sources on or near the WD cooling curve (Fig. \ref{cmd}) which have $UV$ counterparts, and there may also be a few WDs in the 23 sources with $FUV < 22$ but were not detected in the $UV$. 

The other sources in the gap between the WD cooling curve and the MS are expected to include a number of CVs. A GC such as M30 should produce $\approx 200$ CVs by 13 Gyr \citep{ivanova}, through various formation channels such as primordial CVs in which the binary components are the original two stars, and also dynamical binary exchange encounters and tidal capture. However, in the dense core, interactions with other cluster members mean that the rate of destruction of CVs is greater than the formation rate \citep{belloni}, and CVs are expected to have shorter lifetimes than their field counterparts by a factor of three \citep{Shara_2006}. \citet{belloni} find that around 45\% of detectable CVs are within the half-light radius of a GC, which for M30 is $1.03'$ \citep{harris}. After scaling down to our field of view, and taking the destruction rate into account, we have a predicted number of detectable CVs in our images of $\approx 8$. The remaining sources in this region of the CMD that are not CVs or individual WDs are likely to be detached WD-MS binaries, where the hot emission from the WD surface and the cooler radiation from the MS surface places the combined flux of the binary in the region between these two populations. There is also the possibility of chance superpositions, and following the same prescription that was used for the whole dataset in Sect. \ref{transformation}, the expected number of spurious matches in the WD/Gap population is $\approx 1$. Additionally, some sources which lie close to the MS and BS sequence may indeed be MS or BS stars. We stress that the predicted number of CVs is a rough estimate, and a closer investigation into the particular sources that are CV candidates is given in the following publication in this study.
 
We note that for the remainder of this study, most of the sources discussed have been determined as belonging to certain populations by their location in the $FUV - UV$ CMD. Yet there may be some sources which truly belong to a different group, as mentioned in the preceding paragraph of the gap sources close to the MS and BS sequence. This is especially true for the sources near the MS turnoff that have been included in the BS group for example, but may really be MS stars or gap objects, and vice versa. These initial distinctions allow us to make useful preliminary investigations into the cluster population, however supplementary tools such as spectroscopy and additional multi-wavelength observations would enable us to more accurately determine a star's true nature. 

\begin{figure}
    
    \includegraphics[scale=0.07]{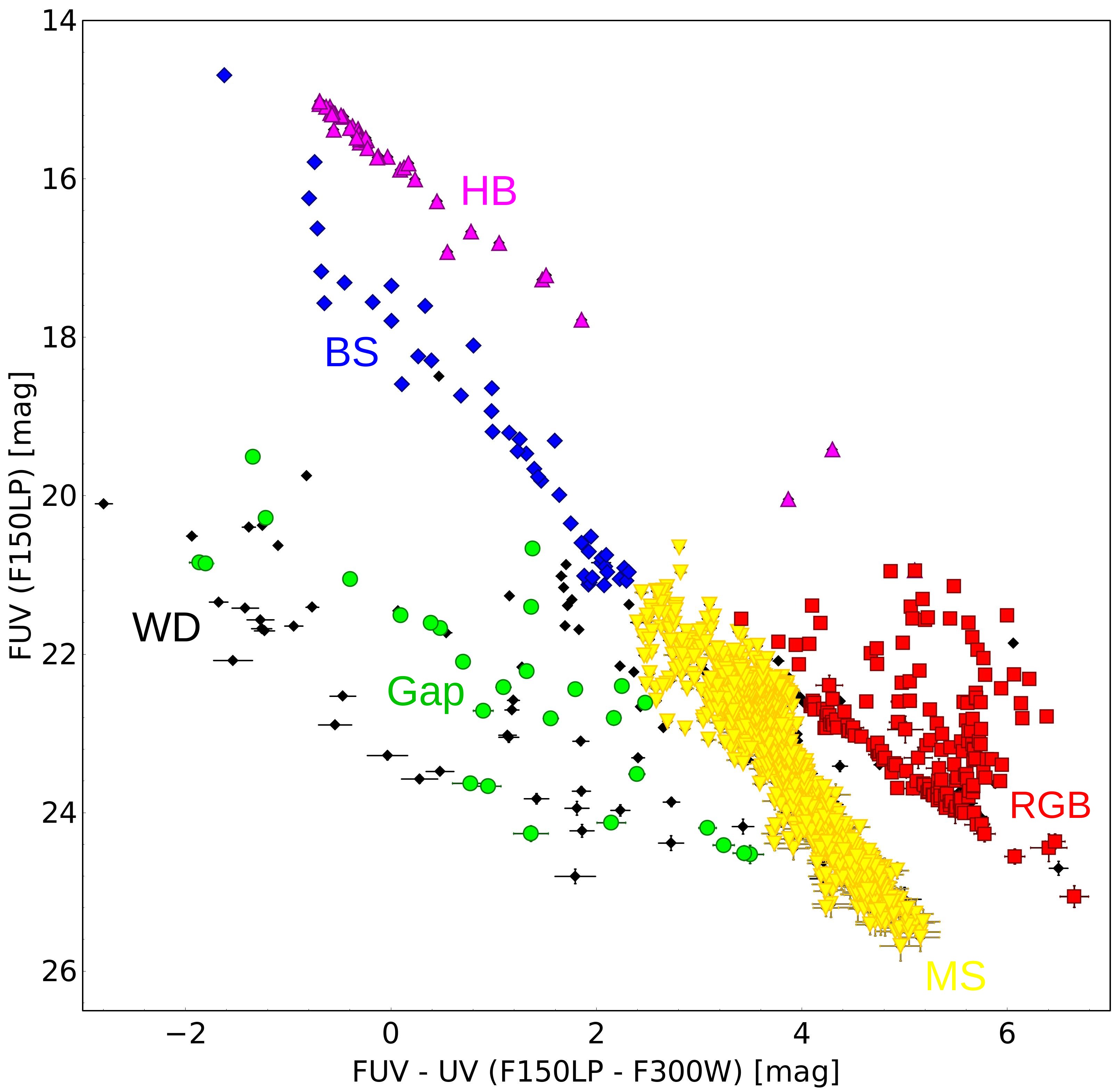}
    
    \includegraphics[scale=0.071]{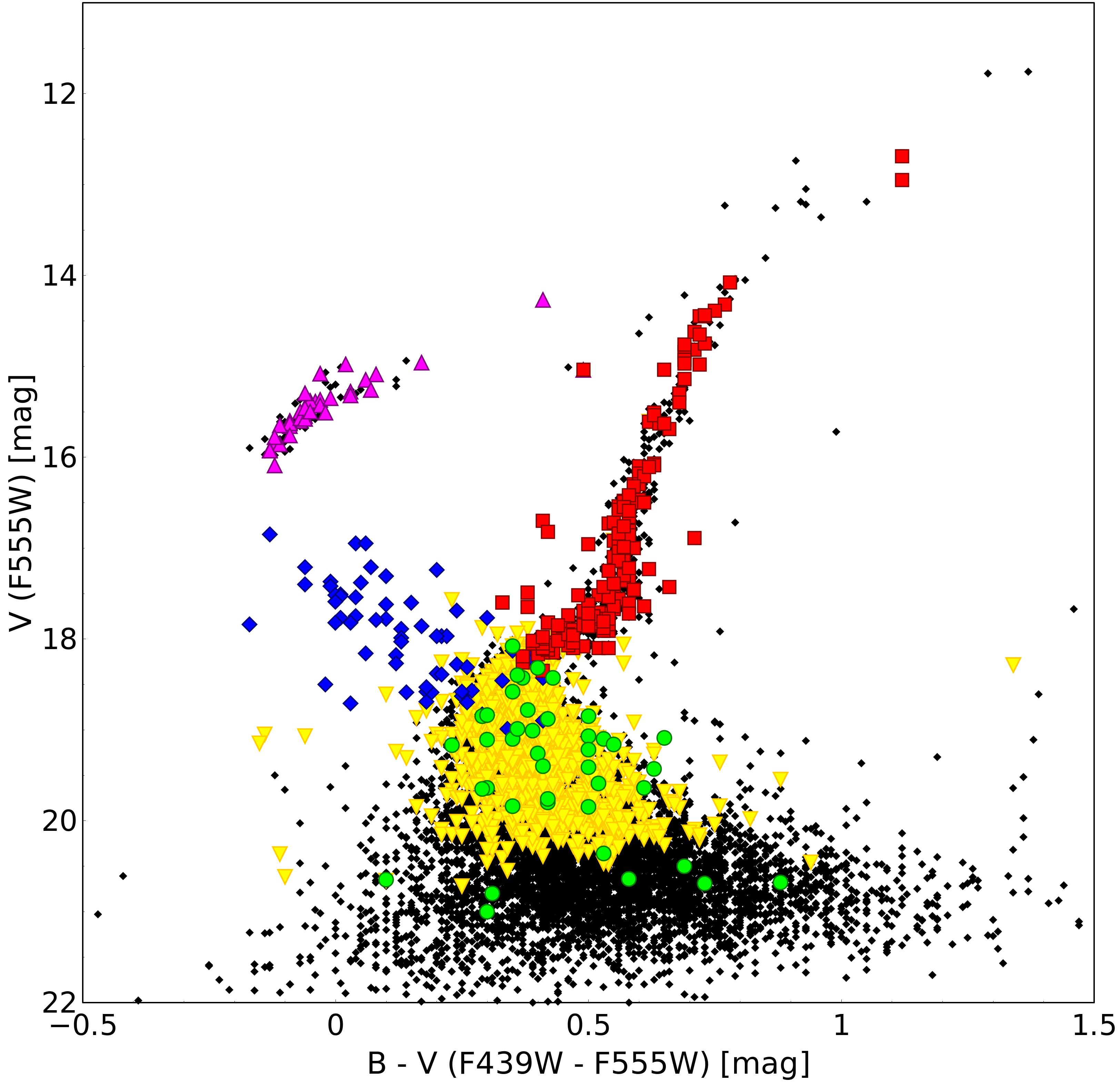}
    \caption{Comparison of the various stellar populations in the \textit{FUV} -- \textit{UV} CMD (top) and the optical CMD (bottom, based on the \citealt{guhathakurta} catalogue). Only the sources present in both catalogues are marked in colour. Included are MS stars (yellow down triangles), red giants (red squares), HB stars (purple up triangles), BS stars (blue diamonds), and gap objects (which include CV candidates, green circles).}
    \label{optical}
\end{figure}

\section{Comparison to optical}\label{opticalsec}

\vspace{5pt}
\citet{guhathakurta} presented an optical catalogue of 9507 sources for M30, the exposures of which were also taken with the WFPC2 on board the $HST$ (program GO-5324, PI: Yanny). Eight exposures were taken on 31st March 1994, two using the F336W filter with an 100 s exposure time, two using the F439W filter and 40\:s exposure time and four using the F555W filter with 4 s exposure times. The data was accessed from VizeiR\footnote{\url{https://cdsarc.cds.unistra.fr/viz-bin/cat/J/AJ/116/1757}}, and by taking the well-defined group of HB stars in the optical CMD (Fig. \ref{optical}), and over-plotting their coordinates onto the \textit{FUV} image, it is easily seen that they corresponded to the brightest sources in the \textit{FUV} image, albeit with a slight shift. After correcting for the shift, these are then used as reference stars to convert the coordinates of the optical catalogue into our \textit{FUV} frame. The coordinates of the HB stars are given as input for the task \texttt{GEOMAP} which computes the transformation. The \texttt{GEOXYTRAN} task carries out the transformation of the optical catalogue into the \textit{FUV} frame, and then also into the \textit{UV} frame using the same transformation used in Sect. \ref{transformation}. The optical coordinates are first converted into the \textit{FUV} frame rather than directly into the \textit{UV} frame, because it was easy to identify stars to use for the transformation, as the HB stars are the brightest objects in the \textit{FUV} and as such are an efficient choice for the reference stars.

The resulting comparison is given in Fig. \ref{optical}. Groups of sources belonging to both catalogues are plotted in colour. There are 1201 matching MS stars (yellow), 178 RGB stars (red), 44 BS sources (blue), all 41 HB stars (purple), and 42 WD/Gap objects (green) which include CV candidates. The comparison illustrates the importance of far-ultraviolet observations in detecting and identifying stellar exotica such as WDs and CVs, as in the optical they are indistinguishable in position from the MS, but their hot emission in the $FUV$ shifts their location in the $FUV - UV$ CMD blueward and brighter than the MS.  This is also true for a few \textit{FUV}-bright BS candidates that are in the MS region in the optical CMD. These sources appear brighter than the MS turnoff in the \textit{FUV} suggesting that they could have a hot WD companion \citep{sahu}. Additionally, two of our BS sources, ID7 and ID66, do not have an optical counterpart. Source ID7 was just out of the field of view in the optical exposures, and ID66 is close to the MS turnoff in the $FUV - UV$ CMD, and as such it may be a binary system with a hot WD that is bright in the ultraviolet but with an optically faint MS star companion \citep{47tuc}. If this is the case, it would really be a Gap object rather than a BS star, as discussed at the end of Sect. \ref{cmd}.

Several sources at the top of the optical RGB do not have \textit{FUV} counterparts as they are located outside the \textit{FUV} field of view. Additionally there were 76 sources in the \textit{FUV} that are undetected in the optical, including nearly half of the WD/Gap objects. This is expected for the sources in the WD cooling region as WDs are optically very faint. The reason for the Gap objects not having optical counterparts may be due to them being CVs with very low-mass MS companions, as their radial distributions suggest, and as such were too faint to be detected in the optical. The MS stars given in the optical catalogue fainter than V $\approx$ 20.5 mag are too faint to be detected in the \textit{FUV}. The matched optical counterparts are identified in Table \ref{fullcat}.

\section{Radial distribution}\label{radialsec}

\begin{figure}
    \centering
    \vspace*{-7pt}
    \includegraphics[width=\columnwidth]{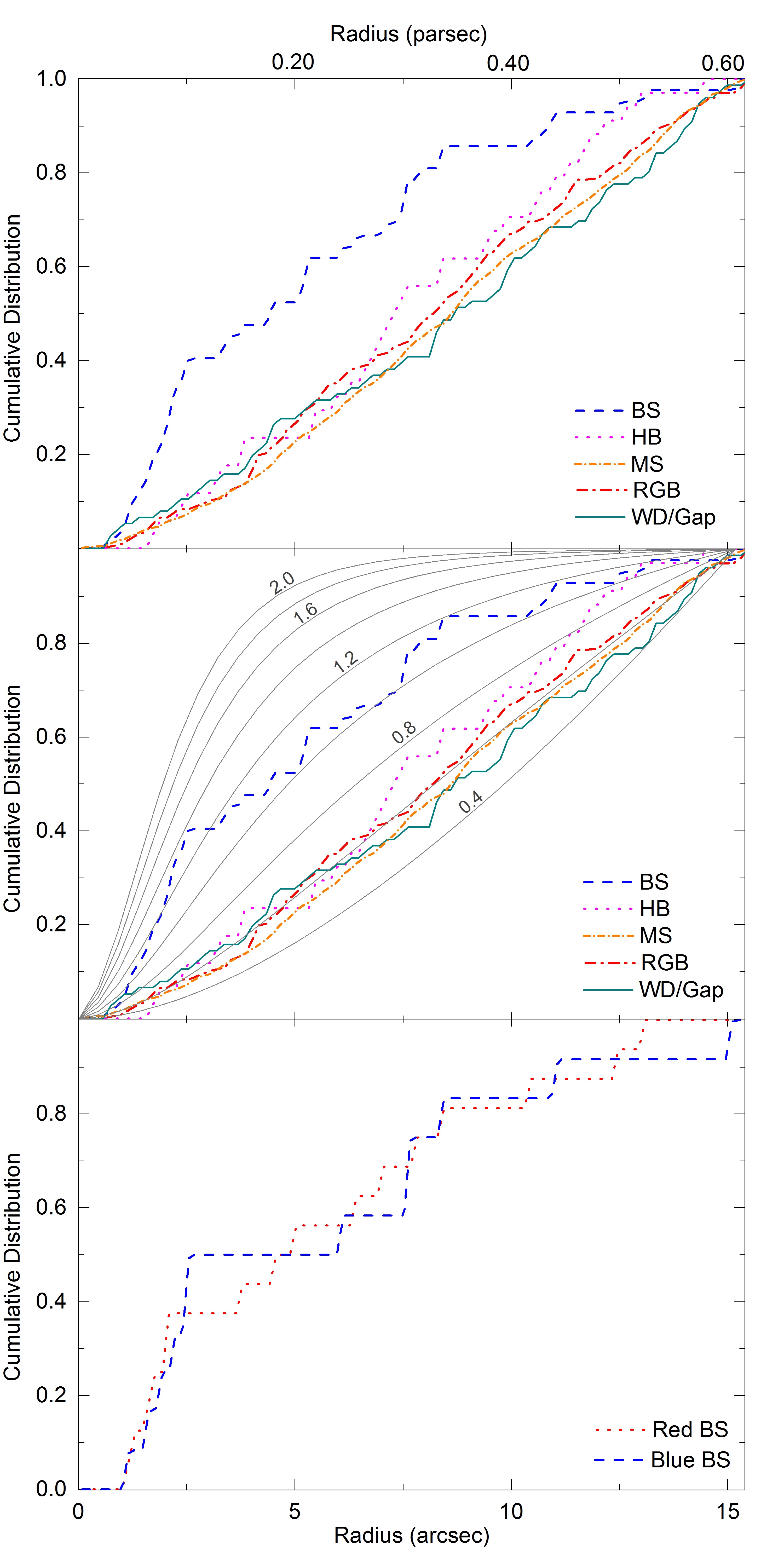}
    \caption{Cumulative radial distributions of the various stellar populations (top) with mass estimate models ranging from 0.4 $M_{\odot}$ -- 2.0 $M_{\odot}$ (middle), and the sources on the red and blue BS sequences (bottom) found by \citet{ferraro}, within a radius of 15.5$''$ from the centre of M30.}
    
    \label{radial}
\end{figure}

\vspace{5pt}
Figure \ref{radial} shows the cumulative radial distributions of the stellar populations found in both the \textit{FUV} and \textit{UV} images within a radius of 15.5$''$ from the cluster centre. The centre of the cluster used was   $\alpha=21^h40^m22\fs13,\ \delta=-23^{\circ}10'47\farcs40$ \citep{ferraro} which corresponds to pixel coordinates $x$ = 443.9, $y$ = 362.8 in the \textit{UV} frame and  $x$ = 612.1, $y$ = 800.1 in the $FUV$.
The numbers for each stellar population in the \textit{FUV}\:--\:\textit{UV} catalogue and those within 15.5$''$ radius of the cluster centre are given in Table \ref{radialtable}. Included in the group WD/Gap sources are all sources blueward of the MS/BS sequence and \textit{FUV} $>$ 19 mag.

The most prominent result in the radial distributions is the strong concentration of BS stars towards the centre, compared to the other populations. M30 is a core-collapsed cluster, meaning that over time the more massive stars have concentrated inward towards the centre (mass segregation). BS stars are the most massive stars out of these stellar populations, having gained mass by either mass transfer from a companion or by the merger of two stars \citep{mccrea, hills}, and as they are still hydrogen burning they have not yet undergone the mass loss observed in the later stages of evolution. They are also likely to form in the cluster centre due to the high number of stellar interactions in this dense region. Thus the BS concentration towards the centre is expected evidence of the dynamical history of the core, and is also representative of the bluer inward colour gradient observed in M30 \citep{howell}. 

\begin{figure}
    \centering
    \includegraphics[width=\columnwidth]{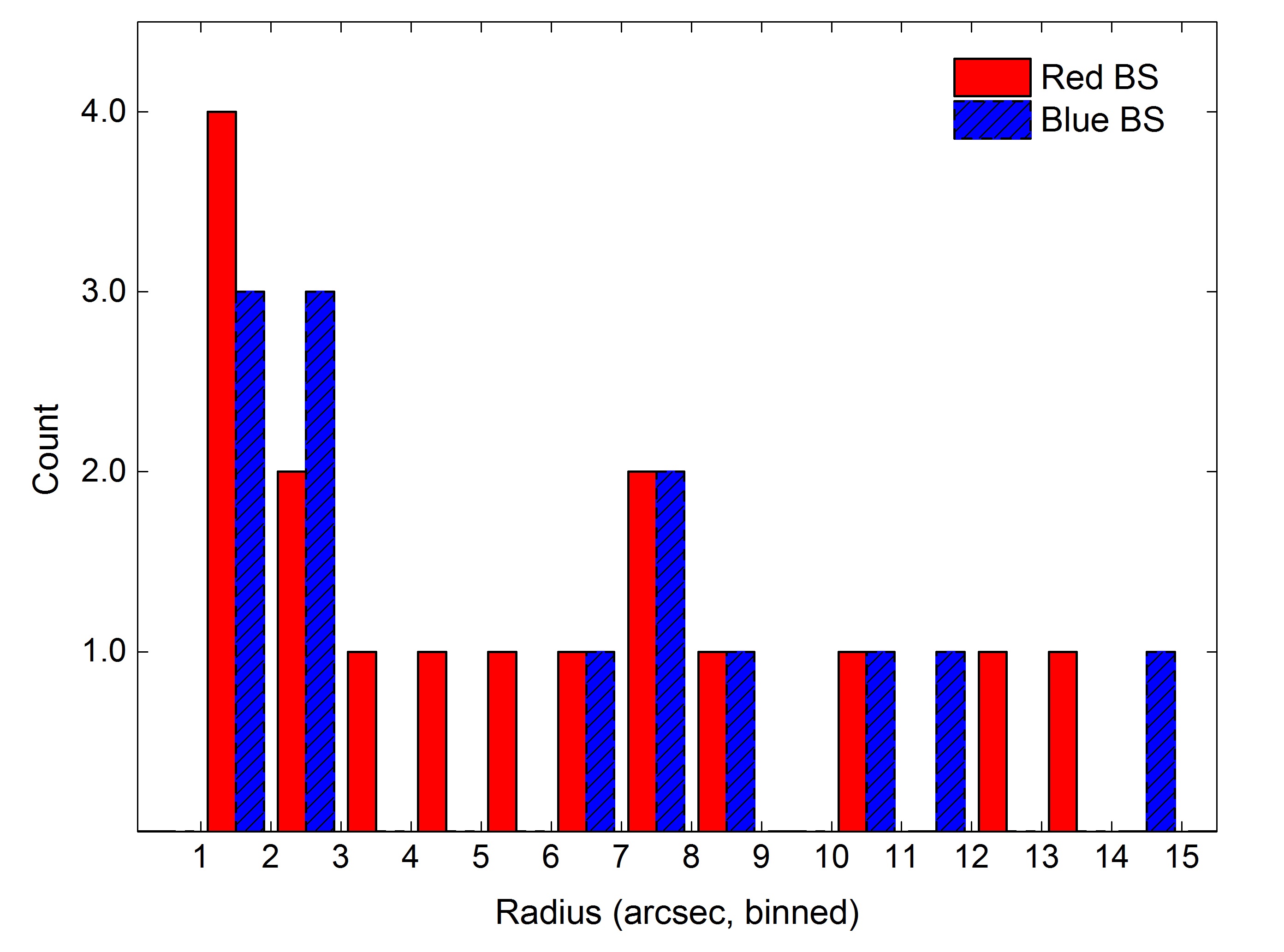}
    \caption{Discrete radial distribution of the red and blue BS sources. The radius is split into 1 arcsec-sized bins, where any red (blue) BS sources appear on the left (right) side of each bin.}
    \label{radBS}
\end{figure}

The colour gradient also results from the central deficiency of RGB stars \citep{howell}, which corresponds to the lack of central concentration of this population shown in Fig. \ref{radial}. This may be due to the mass that stars typically lose during the RGB phase which could allow these stars to drift outwards. This may also be the case for the least centrally concentrated population, the HB stars, which evolve from the RGB stage. The HB stars are even absent within the inner core region of 1.9$''$ (0.08 pc), so whilst the high concentration of BS stars in the centre may add to the bluer inward colour gradient of M30, it seems that the numerous HB stars, which are similar to BS in colour magnitude, do not contribute to this gradient.

The group WD/Gap sources includes $\approx 20$ sources that are likely isolated WD stars that are located near the WD cooling curve (Fig. \ref{isochrones}), an estimated number of $\approx$ 8 CVs, and detached WD-MS binaries. Overall, the sources in this group are only slightly more centrally located than MS stars in M30, which is unexpected at first glance, since a central concentration was seen in other GCs such as NGC\:2808 \citep{ngc2808} and M15 \citep{m15}, and as CVs can form from two-body interactions that are frequent in the high density centre of the cluster. CVs are binary systems, which along with the non-interacting WD-MS binaries, means a higher combined mass than the individual stars alone. Similarly to BS stars, these higher mass sources should concentrate towards the cluster centre over time. However, the segregation of WD/Gap sources towards the centre is not observed, and a possible explanation for this might be that the combined mass of these binary systems is actually relatively low. This is true for old ($\approx$ 10 Gyr) CVs where the WD has devoured almost all of the mass from their companion, leaving behind as little as $\approx\:0.1\:\textnormal{M}_{\odot}$, a negligible mass relative to the WD (\citealt{hillman}), and also mass is lost from the system over time due to wind and outbursts from the accretion disc \citep{tout1991}. Additionally, as the destruction rate of CVs is higher than the formation rate in the dense cores of GCs \citep{belloni2018b}, a number of the detached WD-MS binaries may also have undergone mass transfer in the past. Even if we also detect younger systems, it is likely that the MS companion has a low mass as otherwise the flux from the MS star would be high at the UV wavelength which is not the case for our Gap sources. As there is no central concentration seen in M30, this suggests that a large proportion of the Gap sources are binary systems with low combined masses.

\begin{table}
    \centering
        \caption{Numbers of sources for the stellar populations in the \textit{FUV} -- \textit{UV} catalogue and within a radius of 15.5$''$ of the cluster centre of gravity, along with their estimated masses.}
    \begin{tabular}{cccc}
    \hline
Population & Catalogue & 15.5$''$ radius&Mass\\ \hline\hline
MS         & 1218      & 1030 & 0.5 - 0.6 $M_{\odot}$\\
BS         & 47        & 42 & 1.0 - 1.2 $M_{\odot}$\\
RG         & 185       & 169 & 0.6 - 0.7 $M_{\odot}$\\
HB         & 41        & 34 & 0.6 - 0.8 $M_{\odot}$\\
WD/Gap    & 78      & 70 & 0.5 - 0.6 $M_{\odot}$\\
\hline
    \end{tabular}
    \label{radialtable}
\end{table}

The average masses of the various stellar populations can be estimated from the radial distributions by assuming the spatial distribution of a typical star is described by \citet{king} models. Following the method by \citet{heinke}, we compare our radial distributions using a maximum-likelihood fitting to generalised theoretical King models, with the radial profile of the source surface density:

\begin{equation}
    S(r) = \bigintss \left(1 + \left(\frac{r}{r_{c\star}}\right)^2 \right)^{\frac{1-3q}{2}} dr
\end{equation}

\vspace{5pt}
\noindent where $q = M_X/M_{\star}$, with $M_X$ being the mass of the source population we wish to find and $M_{\star}$ is the mass of the population within the core radius $r_{c\star} = 1.9''$ \citep{sosin}. After applying a correction to the distribution to cover the non-circular field of view of the images as a function of radius, models with masses $0.4 - 2.0\:M_{\odot}$ in steps of 0.2 $M_{\odot}$ are calculated and shown in Fig. \ref{radial}. The estimated masses for each stellar population are given in Table \ref{radialtable}. The average mass of the BS stars is estimated to be $1.0 - 1.2\:M_{\odot}$ and the mass of the WD/Gap sources is $0.5 - 0.6\:M_{\odot}$. Whilst only a slight overall concentration of WD/Gap sources towards the centre was seen, the sources near to the core are estimated to be more massive than those further out. 

The radial distribution of the red and blue BS sources identified as two distinct sequences in \citet{ferraro} is also shown in Fig. \ref{radial} which reveals no significant distinction in radial spread for the two BS sequences; both populations are centrally concentrated. \citet{ferraro} show a slightly higher central concentration of red BS in M30, although they use a different cumulative radial range, therefore we also give the discrete red and blue BS radial distributions in Fig.\:\ref{radBS}, again displaying no significant difference between the two sub-populations. From the other two clusters known to have a double BS sequence, \citet{Dalessandro_2013} find a higher central concentration of red BSs in NGC 362 and contrastingly, \citet{Simunovic_2014} find a higher central concentration of blue BS stars in NGC 1261.

The similarity of the radial distributions of the various stellar populations is investigated using  
a Kolmogorov--Smirnov (KS) test. The KS test compares two populations and returns a probability that the two distributions are drawn from the same parent population. The higher the returned probability, the more likely the distributions are from the same population. The BS sources are comparable to the other stellar populations with KS probabilities of 0.54$\%$ (BS to HB), 0.05$\%$ (BS to RG), and 0.02$\%$ (BS to WD/Gap sources), suggesting that the BS sources formed through a different process to the HB and RG stars, i.e. from mergers or mass transfer rather than single star evolution. The results for all population comparisons are given in Table \ref{KS}.

\begin{table}
    \centering
\caption{Probabilities that two populations of stars are from the same parent population.}
    \begin{tabular}{lc}
    
\hline

Population&Probabilities\\ 
Comparison&\%\\
\hline\hline
MS--BS &\hspace{5pt}0.001\\
MS--HB &47.996\\
MS--RG &40.998\\
MS--WD/Gap &62.156\\
BS--HB &0.5416\\
BS--RG &0.0452\\
BS--WD/Gap &0.0201\\
HB--RG &78.936\\
HB--WD/Gap &36.181\\
RG--WD/Gap &53.490\\
\hline
    \end{tabular}

    \label{KS}
\end{table}

\section{Summary} \label{summ}

\vspace{5pt}
Far-ultraviolet (\textit{FUV}, ACS/SBC/F150LP) and mid-ultraviolet (\textit{UV}, WFPC2/F300W) exposures of the globular cluster M30 were
photometrically analysed. A total of 1934 sources are detected in the \textit{FUV} image and 10451 sources in the \textit{UV} image. Out of these, 1569 matching sources were found. Different stellar populations are well distinguished in the resulting \textit{FUV} $-$ \textit{UV} CMD. The MS turnoff lies at $\textnormal{\textit{FUV}} \approx 22$ mag and $\textnormal{\textit{FUV} -- \textit{UV}} \approx 3$ mag. The horizontal branch consists of 41 sources, all of which have an optical counterpart from the catalogue by \citet{guhathakurta}. The red giant branch extends towards fainter and redder magnitudes, with 185 RGB sources, 178 of which are also found in the optical. A sequence of 47 BS stars is observed, 44 of these having optical counterparts. Seventy-eight WD/Gap objects are identified, 42 of which were in the optical catalogue. The $FUV - UV$ CMD allows us to easily distinguish the WD/Gap sources from MS stars.

The double BS sequence suggested by \citet{ferraro} in the $V - I$ CMD is not seen in the \textit{FUV} $-$ \textit{UV} CMD. The two sets of BS sources are mixed in the ultraviolet which may be a result of the sources on the red BS sequence experiencing active mass transfer and emitting $FUV$ radiation, shifting these sources blueward in the ultraviolet CMD.

The radial distributions of the stellar populations show a strong concentration of BS sources towards the centre of the cluster, implying that mass segregation has taken place. There is a deficiency of HB stars in the very centre of the cluster and no central concentration of CV
candidates is found which may be due to these being old systems with low-mass MS companions.

The \textit{HST} data for this project is sensitive enough to detect a previously unseen, significant population of WD/Gap objects, providing new insight into M30 in the ultraviolet. Investigations in the \textit{FUV} continue to provide detailed detections and identifications of their stellar populations.

\section*{Data Availability}

\vspace{5pt}
The data underlying this article are available in the Mikulski Archive for Space Telescopes (MAST): \url{https://archive.stsci.edu}. The datasets are derived from images in the public domain: \url{https://archive.stsci.edu/proposal\_search.php?mission=hst&id=10561}. The catalogue of sources is available at CDS via anonymous ftp to cdsarc.u-strasbg.fr (130.79.128.5) or via \url{https://cdsarc.unistra.fr/viz-bin/cat/J/MNRAS}. This research also made use of the optical data available at CDS: \url{https://cdsarc.cds.unistra.fr/viz-bin/cat/J/AJ/116/1757}.

\section*{Acknowledgements}

\vspace{5pt}
PK acknowledges support from the Grant Agency of the Czech Republic under grant number 20-21855S.

\raggedright
\bibliographystyle{mnras}
\bibliography{M30.bib} 

\begin{thebibliography}{}
\makeatletter
\relax
\def\mn@urlcharsother{\let\do\@makeother \do\$\do\&\do\#\do\^\do\_\do\%\do\~}
\def\mn@doi{\begingroup\mn@urlcharsother \@ifnextchar [ {\mn@doi@}
  {\mn@doi@[]}}
\def\mn@doi@[#1]#2{\def\@tempa{#1}\ifx\@tempa\@empty \href
  {http://dx.doi.org/#2} {doi:#2}\else \href {http://dx.doi.org/#2} {#1}\fi
  \endgroup}
\def\mn@eprint#1#2{\mn@eprint@#1:#2::\@nil}
\def\mn@eprint@arXiv#1{\href {http://arxiv.org/abs/#1} {{\tt arXiv:#1}}}
\def\mn@eprint@dblp#1{\href {http://dblp.uni-trier.de/rec/bibtex/#1.xml}
  {dblp:#1}}
\def\mn@eprint@#1:#2:#3:#4\@nil{\def\@tempa {#1}\def\@tempb {#2}\def\@tempc
  {#3}\ifx \@tempc \@empty \let \@tempc \@tempb \let \@tempb \@tempa \fi \ifx
  \@tempb \@empty \def\@tempb {arXiv}\fi \@ifundefined
  {mn@eprint@\@tempb}{\@tempb:\@tempc}{\expandafter \expandafter \csname
  mn@eprint@\@tempb\endcsname \expandafter{\@tempc}}}

\bibitem[\protect\citeauthoryear{{Allen}, {Moreno}  \& {Pichardo}}{{Allen}
  et~al.}{2006}]{allen2006}
{Allen} C.,  {Moreno} E.,   {Pichardo} B.,  2006, \mn@doi [\apj]
  {10.1086/508676}, \href
  {https://ui.adsabs.harvard.edu/abs/2006ApJ...652.1150A} {652, 1150}

\bibitem[\protect\citeauthoryear{{Althaus} \& {Benvenuto}}{{Althaus} \&
  {Benvenuto}}{1998}]{althaus}
{Althaus} L.~G.,  {Benvenuto} O.~G.,  1998, \mn@doi [\mnras]
  {10.1046/j.1365-8711.1998.01332.x}, \href
  {https://ui.adsabs.harvard.edu/abs/1998MNRAS.296..206A} {296, 206}

\bibitem[\protect\citeauthoryear{{Avila} \& {Chiaberge}}{{Avila} \&
  {Chiaberge}}{2016}]{acs}
{Avila} R.~J.,  {Chiaberge} M.,  2016, STScI Instrument Science Report ACS
  2016-5, \href {https://ui.adsabs.harvard.edu/abs/2016acs..rept....5A} {}

\bibitem[\protect\citeauthoryear{Belloni, Zorotovic, Schreiber, Leigh, Giersz
  \& Askar}{Belloni et~al.}{2017}]{belloni2017}
Belloni D.,  Zorotovic M.,  Schreiber M.~R.,  Leigh N. W.~C.,  Giersz M.,
  Askar A.,  2017, \mn@doi [\mnras] {10.1093/mnras/stx575}, 468, 2429

\bibitem[\protect\citeauthoryear{Belloni, Schreiber, Zorotovic, Iłkiewicz,
  Hurley, Giersz  \& Lagos}{Belloni et~al.}{2018a}]{belloni2018a}
Belloni D.,  Schreiber M.~R.,  Zorotovic M.,  Iłkiewicz K.,  Hurley J.~R.,
  Giersz M.,   Lagos F.,  2018a, \mn@doi [\mnras] {10.1093/mnras/sty1421}, 478,
  5626

\bibitem[\protect\citeauthoryear{Belloni, Giersz, Rivera Sandoval, Askar  \&
  Ciecieląg}{Belloni et~al.}{2018b}]{belloni2018b}
Belloni D.,  Giersz M.,  Rivera Sandoval L.~E.,  Askar A.,   Ciecieląg P.,
  2018b, \mn@doi [\mnras] {10.1093/mnras/sty3097}, 483, 315

\bibitem[\protect\citeauthoryear{{Belloni}, {Giersz}, {Sandoval}, {Askar}  \&
  {Ciecielag}}{{Belloni} et~al.}{2020}]{Belloni2019}
{Belloni} D.,  {Giersz} M.,  {Sandoval} L. E.~R.,  {Askar} A.,   {Ciecielag}
  P.,  2020, \mn@doi [Proc. IAU] {10.1017/S174392131900718X}, \href
  {https://ui.adsabs.harvard.edu/abs/2020IAUS..351..404B} {351, 404}

\bibitem[\protect\citeauthoryear{Breen \& Heggie}{Breen \&
  Heggie}{2013}]{breen}
Breen P.~G.,  Heggie D.~C.,  2013, \mn@doi [MNRAS] {10.1093/mnras/stt1599},
  436, 584

\bibitem[\protect\citeauthoryear{{Brown}, {Sweigart}, {Lanz}, {Landsman}  \&
  {Hubeny}}{{Brown} et~al.}{2001}]{brown}
{Brown} T.~M.,  {Sweigart} A.~V.,  {Lanz} T.,  {Landsman} W.~B.,   {Hubeny} I.,
   2001, \mn@doi [\apj] {10.1086/323862}, \href
  {https://ui.adsabs.harvard.edu/abs/2001ApJ...562..368B} {562, 368}

\bibitem[\protect\citeauthoryear{{Chatterjee}, {Fregeau}, {Umbreit}  \&
  {Rasio}}{{Chatterjee} et~al.}{2010}]{chatterjee2010}
{Chatterjee} S.,  {Fregeau} J.~M.,  {Umbreit} S.,   {Rasio} F.~A.,  2010,
  \mn@doi [\apj] {10.1088/0004-637X/719/1/915}, \href
  {https://ui.adsabs.harvard.edu/abs/2010ApJ...719..915C} {719, 915}

\bibitem[\protect\citeauthoryear{{Chatterjee}, {Rasio}, {Sills}  \&
  {Glebbeek}}{{Chatterjee} et~al.}{2013}]{chatterjee2013}
{Chatterjee} S.,  {Rasio} F.~A.,  {Sills} A.,   {Glebbeek} E.,  2013, \mn@doi
  [\apj] {10.1088/0004-637X/777/2/106}, \href
  {https://ui.adsabs.harvard.edu/abs/2013ApJ...777..106C} {777, 106}

\bibitem[\protect\citeauthoryear{{Dalessandro}, {Beccari}, {Lanzoni},
  {Ferraro}, {Schiavon}  \& {Rood}}{{Dalessandro} et~al.}{2009}]{m2}
{Dalessandro} E.,  {Beccari} G.,  {Lanzoni} B.,  {Ferraro} F.~R.,  {Schiavon}
  R.,   {Rood} R.~T.,  2009, \mn@doi [\apjs] {10.1088/0067-0049/182/2/509},
  \href {https://ui.adsabs.harvard.edu/abs/2009ApJS..182..509D} {182, 509}

\bibitem[\protect\citeauthoryear{Dalessandro, Schiavon, Rood, Ferraro, Sohn,
  Lanzoni  \& O'Connell}{Dalessandro et~al.}{2012}]{Dalessandro2012}
Dalessandro E.,  Schiavon R.~P.,  Rood R.~T.,  Ferraro F.~R.,  Sohn S.~T.,
  Lanzoni B.,   O'Connell R.~W.,  2012, \mn@doi [ApJ]
  {10.1088/0004-6256/144/5/126}, 144, 126

\bibitem[\protect\citeauthoryear{Dalessandro, Salaris, Ferraro, Mucciarelli  \&
  Cassisi}{Dalessandro et~al.}{2013a}]{dalessandro2013}
Dalessandro E.,  Salaris M.,  Ferraro F.~R.,  Mucciarelli A.,   Cassisi S.,
  2013a, \mn@doi [MNRAS] {10.1093/mnras/sts644}, 430, 459

\bibitem[\protect\citeauthoryear{Dalessandro et~al.,}{Dalessandro
  et~al.}{2013b}]{Dalessandro_2013}
Dalessandro E.,  et~al., 2013b, \mn@doi [ApJ] {10.1088/0004-637x/778/2/135},
  778, 135

\bibitem[\protect\citeauthoryear{{Dieball}, {Knigge}, {Zurek}, {Shara}  \&
  {Long}}{{Dieball} et~al.}{2005}]{ngc2808}
{Dieball} A.,  {Knigge} C.,  {Zurek} D.~R.,  {Shara} M.~M.,   {Long} K.~S.,
  2005, \mn@doi [\apj] {10.1086/429534}, \href
  {https://ui.adsabs.harvard.edu/abs/2005ApJ...625..156D} {625, 156}

\bibitem[\protect\citeauthoryear{Dieball, Knigge, Zurek, Shara, Long, Charles
  \& Hannikainen}{Dieball et~al.}{2007}]{m15}
Dieball A.,  Knigge C.,  Zurek D.~R.,  Shara M.~M.,  Long K.~S.,  Charles
  P.~A.,   Hannikainen D.,  2007, \mn@doi [ApJ] {10.1086/522103}, 670, 379

\bibitem[\protect\citeauthoryear{Dieball, Long, Knigge, Thomson  \&
  Zurek}{Dieball et~al.}{2010}]{m80_2}
Dieball A.,  Long K.~S.,  Knigge C.,  Thomson G.~S.,   Zurek D.~R.,  2010,
  \mn@doi [ApJ] {10.1088/0004-637x/710/1/332}, 710, 332

\bibitem[\protect\citeauthoryear{Dieball, Rasekh, Knigge, Shara  \&
  Zurek}{Dieball et~al.}{2017}]{ngc6397}
Dieball A.,  Rasekh A.,  Knigge C.,  Shara M.,   Zurek D.,  2017, \mn@doi
  [MNRAS] {10.1093/mnras/stx802}, 469, 267

\bibitem[\protect\citeauthoryear{{Dolphin}}{{Dolphin}}{2000}]{dolphot}
{Dolphin} A.~E.,  2000, \mn@doi [\pasp] {10.1086/316630}, \href
  {https://ui.adsabs.harvard.edu/abs/2000PASP..112.1383D} {112, 1383}

\bibitem[\protect\citeauthoryear{{Dorman}}{{Dorman}}{1992a}]{dorman}
{Dorman} B.,  1992a, \mn@doi [\apjs] {10.1086/191678}, \href
  {https://ui.adsabs.harvard.edu/abs/1992ApJS...80..701D} {80, 701}

\bibitem[\protect\citeauthoryear{{Dorman}}{{Dorman}}{1992b}]{dorman1992a}
{Dorman} B.,  1992b, \mn@doi [\apjs] {10.1086/191691}, \href
  {https://ui.adsabs.harvard.edu/abs/1992ApJS...81..221D} {81, 221}

\bibitem[\protect\citeauthoryear{Ferraro et~al.}{Ferraro
  et~al.}{2009}]{ferraro}
Ferraro F.,  et~al., 2009, \mn@doi [Nature] {10.1038/nature08607}, 462, 1028

\bibitem[\protect\citeauthoryear{{G{\"a}nsicke}, {Beuermann}  \& {de
  Martino}}{{G{\"a}nsicke} et~al.}{1995}]{gansicke}
{G{\"a}nsicke} B.~T.,  {Beuermann} K.,   {de Martino} D.,  1995, \aap, \href
  {https://ui.adsabs.harvard.edu/abs/1995A&A...303..127G} {303, 127}

\bibitem[\protect\citeauthoryear{{Greenfield} \& {White}}{{Greenfield} \&
  {White}}{2000}]{pyraf}
{Greenfield} P.,  {White} R.~L.,  2000, ASPCS, \href
  {https://ui.adsabs.harvard.edu/abs/2000ASPC..216...59G} {216, 59}

\bibitem[\protect\citeauthoryear{{Guhathakurta}, {Webster}, {Yanny},
  {Schneider}  \& {Bahcall}}{{Guhathakurta} et~al.}{1998}]{guhathakurta}
{Guhathakurta} P.,  {Webster} Z.~T.,  {Yanny} B.,  {Schneider} D.~P.,
  {Bahcall} J.~N.,  1998, \mn@doi [\aj] {10.1086/300566}, \href
  {https://ui.adsabs.harvard.edu/abs/1998AJ....116.1757G} {116, 1757}

\bibitem[\protect\citeauthoryear{{Harris}}{{Harris}}{1996}]{harris}
{Harris} W.~E.,  1996, \mn@doi [\aj] {10.1086/118116}, \href
  {https://ui.adsabs.harvard.edu/abs/1996AJ....112.1487H} {112, 1487}

\bibitem[\protect\citeauthoryear{{Heggie}}{{Heggie}}{1975}]{heggie}
{Heggie} D.~C.,  1975, \mn@doi [\mnras] {10.1093/mnras/173.3.729}, \href
  {https://ui.adsabs.harvard.edu/abs/1975MNRAS.173..729H} {173, 729}

\bibitem[\protect\citeauthoryear{Heinke, Grindlay, Edmonds, Lloyd, Murray, Cohn
   \& Lugger}{Heinke et~al.}{2003}]{heinke}
Heinke C.~O.,  Grindlay J.~E.,  Edmonds P.~D.,  Lloyd D.~A.,  Murray S.~S.,
  Cohn H.~N.,   Lugger P.~M.,  2003, \mn@doi [ApJ] {10.1086/378884}, 598, 516

\bibitem[\protect\citeauthoryear{Hillman, Shara, Prialnik  \& Kovetz}{Hillman
  et~al.}{2020}]{hillman}
Hillman Y.,  Shara M.~M.,  Prialnik D.,   Kovetz A.,  2020, \mn@doi [Nat
  Astron] {10.1038/s41550-020-1062-y}, 4, 886–892

\bibitem[\protect\citeauthoryear{{Hills}}{{Hills}}{1975}]{hills1975}
{Hills} J.~G.,  1975, \mn@doi [AJ] {10.1086/111815}, \href
  {https://ui.adsabs.harvard.edu/abs/1975AJ.....80..809H} {80, 809}

\bibitem[\protect\citeauthoryear{{Hills} \& {Day}}{{Hills} \&
  {Day}}{1976}]{hills}
{Hills} J.~G.,  {Day} C.~A.,  1976, \aplett, \href
  {https://ui.adsabs.harvard.edu/abs/1976ApL....17...87H} {17, 87}

\bibitem[\protect\citeauthoryear{Hong, Vesperini, Belloni  \& Giersz}{Hong
  et~al.}{2016}]{hong}
Hong J.,  Vesperini E.,  Belloni D.,   Giersz M.,  2016, \mn@doi [\mnras]
  {10.1093/mnras/stw2595}, 464, 2511

\bibitem[\protect\citeauthoryear{Howell, Guhathakurta  \& Tan}{Howell
  et~al.}{2000}]{howell}
Howell J.~H.,  Guhathakurta P.,   Tan A.,  2000, \mn@doi [AJ] {10.1086/301270},
  119, 1259

\bibitem[\protect\citeauthoryear{Hurley \& Shara}{Hurley \&
  Shara}{2012}]{hurley_shara2012}
Hurley J.~R.,  Shara M.~M.,  2012, \mn@doi [Monthly Notices of the Royal
  Astronomical Society] {10.1111/j.1365-2966.2012.21668.x}, 425, 2872

\bibitem[\protect\citeauthoryear{Hurley, Aarseth  \& Shara}{Hurley
  et~al.}{2007}]{Hurley_2007}
Hurley J.~R.,  Aarseth S.~J.,   Shara M.~M.,  2007, \mn@doi [ApJ]
  {10.1086/517879}, 665, 707

\bibitem[\protect\citeauthoryear{{Hut} et~al.,}{{Hut} et~al.}{1992}]{hut}
{Hut} P.,  et~al., 1992, \mn@doi [PASP] {10.1086/133085}, \href
  {https://ui.adsabs.harvard.edu/abs/1992PASP..104..981H} {104, 981}

\bibitem[\protect\citeauthoryear{Hypki \& Giersz}{Hypki \&
  Giersz}{2012}]{hypki}
Hypki A.,  Giersz M.,  2012, \mn@doi [MNRAS] {10.1093/mnras/sts415}, 429, 1221

\bibitem[\protect\citeauthoryear{Ivanova, Heinke, Rasio, Taam, Belczynski  \&
  Fregeau}{Ivanova et~al.}{2006}]{ivanova}
Ivanova N.,  Heinke C.~O.,  Rasio F.~A.,  Taam R.~E.,  Belczynski K.,   Fregeau
  J.,  2006, \mn@doi [MNRAS] {10.1111/j.1365-2966.2006.10876.x}, 372, 1043

\bibitem[\protect\citeauthoryear{{Kains, N.} et~al.,}{{Kains, N.}
  et~al.}{2013}]{kains}
{Kains, N.} et~al., 2013, \mn@doi [A\&A] {10.1051/0004-6361/201321819}, 555,
  A36

\bibitem[\protect\citeauthoryear{{King}}{{King}}{1966}]{king}
{King} I.~R.,  1966, \mn@doi [\aj] {10.1086/109857}, \href
  {https://ui.adsabs.harvard.edu/abs/1966AJ.....71...64K} {71, 64}

\bibitem[\protect\citeauthoryear{Knigge, Shara, Zurek, Long  \&
  Gilliland}{Knigge et~al.}{2000}]{47tuc}
Knigge C.,  Shara M.~M.,  Zurek D.~R.,  Long K.~S.,   Gilliland R.~L.,  2000,
  ASP Conference Series (\mn@eprint {} {astro-ph/0012187})

\bibitem[\protect\citeauthoryear{Knigge, Zurek, Shara, Long  \&
  Gilliland}{Knigge et~al.}{2003}]{knigge}
Knigge C.,  Zurek D.~R.,  Shara M.~M.,  Long K.~S.,   Gilliland R.~L.,  2003,
  \mn@doi [ApJ] {10.1086/379609}, 599, 1320

\bibitem[\protect\citeauthoryear{{Kremer} et~al.,}{{Kremer}
  et~al.}{2020}]{kremer2020}
{Kremer} K.,  et~al., 2020, \mn@doi [\apjs] {10.3847/1538-4365/ab7919}, \href
  {https://ui.adsabs.harvard.edu/abs/2020ApJS..247...48K} {247, 48}

\bibitem[\protect\citeauthoryear{{Leigh}, {Sills}  \& {Knigge}}{{Leigh}
  et~al.}{2007}]{leigh2007}
{Leigh} N.,  {Sills} A.,   {Knigge} C.,  2007, \mn@doi [\apj] {10.1086/514330},
  \href {https://ui.adsabs.harvard.edu/abs/2007ApJ...661..210L} {661, 210}

\bibitem[\protect\citeauthoryear{{Leigh}, {Sills}  \& {Knigge}}{{Leigh}
  et~al.}{2011}]{leigh2011}
{Leigh} N.,  {Sills} A.,   {Knigge} C.,  2011, \mn@doi [\mnras]
  {10.1111/j.1365-2966.2011.18995.x}, \href
  {https://ui.adsabs.harvard.edu/abs/2011MNRAS.415.3771L} {415, 3771}

\bibitem[\protect\citeauthoryear{{Leigh}, {Knigge}, {Sills}, {Perets},
  {Sarajedini}  \& {Glebbeek}}{{Leigh} et~al.}{2013}]{leigh2013}
{Leigh} N.,  {Knigge} C.,  {Sills} A.,  {Perets} H.~B.,  {Sarajedini} A.,
  {Glebbeek} E.,  2013, \mn@doi [\mnras] {10.1093/mnras/sts085}, \href
  {https://ui.adsabs.harvard.edu/abs/2013MNRAS.428..897L} {428, 897}

\bibitem[\protect\citeauthoryear{{Leigh}, {Giersz}, {Marks}, {Webb}, {Hypki},
  {Heinke}, {Kroupa}  \& {Sills}}{{Leigh} et~al.}{2015}]{leigh2015}
{Leigh} N. W.~C.,  {Giersz} M.,  {Marks} M.,  {Webb} J.~J.,  {Hypki} A.,
  {Heinke} C.~O.,  {Kroupa} P.,   {Sills} A.,  2015, \mn@doi [\mnras]
  {10.1093/mnras/stu2110}, \href
  {https://ui.adsabs.harvard.edu/abs/2015MNRAS.446..226L} {446, 226}

\bibitem[\protect\citeauthoryear{Lugger, Cohn, Heinke, Grindlay  \&
  Edmonds}{Lugger et~al.}{2007}]{lugger}
Lugger P.~M.,  Cohn H.~N.,  Heinke C.~O.,  Grindlay J.~E.,   Edmonds P.~D.,
  2007, \mn@doi [ApJ] {10.1086/507572}, 657, 286

\bibitem[\protect\citeauthoryear{Maccarone, Long, Knigge, Dieball  \&
  Zurek}{Maccarone et~al.}{2010}]{maccarone}
Maccarone T.~J.,  Long K.~S.,  Knigge C.,  Dieball A.,   Zurek D.~R.,  2010,
  \mn@doi [MNRAS] {10.1111/j.1365-2966.2010.16833.x}, 406, 2087

\bibitem[\protect\citeauthoryear{McCrea}{McCrea}{1964}]{mccrea}
McCrea W.~H.,  1964, \mn@doi [MNRAS] {10.1093/mnras/128.2.147}, 128, 147

\bibitem[\protect\citeauthoryear{{Parise} et~al.,}{{Parise}
  et~al.}{1994}]{ngc1851}
{Parise} R.~A.,  et~al., 1994, \mn@doi [\apj] {10.1086/173807}, \href
  {https://ui.adsabs.harvard.edu/abs/1994ApJ...423..305P} {423, 305}

\bibitem[\protect\citeauthoryear{{Portegies Zwart}}{{Portegies
  Zwart}}{2019}]{zwart}
{Portegies Zwart} S.,  2019, \mn@doi [\aap] {10.1051/0004-6361/201833485},
  \href {https://ui.adsabs.harvard.edu/abs/2019A&A...621L..10P} {621, L10}

\bibitem[\protect\citeauthoryear{Richer et~al.,}{Richer
  et~al.}{1997}]{Richer1997}
Richer H.~B.,  et~al., 1997, \mn@doi [ApJ] {10.1086/304379}, 484, 741

\bibitem[\protect\citeauthoryear{Rodriguez, Morscher, Wang, Chatterjee, Rasio
  \& Spurzem}{Rodriguez et~al.}{2016}]{rodriguez}
Rodriguez C.~L.,  Morscher M.,  Wang L.,  Chatterjee S.,  Rasio F.~A.,
  Spurzem R.,  2016, \mn@doi [\mnras] {10.1093/mnras/stw2121}, 463, 2109

\bibitem[\protect\citeauthoryear{{Rui}, {Kremer}, {Weatherford}, {Chatterjee},
  {Rasio}, {Rodriguez}  \& {Ye}}{{Rui} et~al.}{2021}]{rui2021}
{Rui} N.~Z.,  {Kremer} K.,  {Weatherford} N.~C.,  {Chatterjee} S.,  {Rasio}
  F.~A.,  {Rodriguez} C.~L.,   {Ye} C.~S.,  2021, \mn@doi [\apj]
  {10.3847/1538-4357/abed49}, \href
  {https://ui.adsabs.harvard.edu/abs/2021ApJ...912..102R} {912, 102}

\bibitem[\protect\citeauthoryear{{Sahu} et~al.,}{{Sahu} et~al.}{2019}]{Sahu}
{Sahu} S.,  et~al., 2019, \mn@doi [\apj] {10.3847/1538-4357/ab11d0}, \href
  {https://ui.adsabs.harvard.edu/abs/2019ApJ...876...34S} {876, 34}

\bibitem[\protect\citeauthoryear{Schiavon et~al.,}{Schiavon
  et~al.}{2012}]{Schiavon2012}
Schiavon R.~P.,  et~al., 2012, \mn@doi [ApJ] {10.1088/0004-6256/143/5/121},
  143, 121

\bibitem[\protect\citeauthoryear{Shara \& Hurley}{Shara \&
  Hurley}{2006}]{Shara_2006}
Shara M.~M.,  Hurley J.~R.,  2006, \mn@doi [ApJ] {10.1086/504679}, 646, 464

\bibitem[\protect\citeauthoryear{Siegel et~al.,}{Siegel
  et~al.}{2014}]{Siegel_2014}
Siegel M.~H.,  et~al., 2014, \mn@doi [ApJ] {10.1088/0004-6256/148/6/131}, 148,
  131

\bibitem[\protect\citeauthoryear{Siegel, Porterfield, Balzer  \& Hagen}{Siegel
  et~al.}{2015}]{Siegel_2015}
Siegel M.~H.,  Porterfield B.~L.,  Balzer B.~G.,   Hagen L. M.~Z.,  2015,
  \mn@doi [ApJ] {10.1088/0004-6256/150/4/129}, 150, 129

\bibitem[\protect\citeauthoryear{Sills, Karakas  \& Lattanzio}{Sills
  et~al.}{2009}]{sills}
Sills A.,  Karakas A.,   Lattanzio J.,  2009, \mn@doi [ApJ]
  {10.1088/0004-637x/692/2/1411}, 692, 1411

\bibitem[\protect\citeauthoryear{Simunovic, Puzia  \& Sills}{Simunovic
  et~al.}{2014}]{Simunovic_2014}
Simunovic M.,  Puzia T.~H.,   Sills A.,  2014, \mn@doi [ApJ]
  {10.1088/2041-8205/795/1/l10}, 795, L10

\bibitem[\protect\citeauthoryear{{Sosin}}{{Sosin}}{1997}]{sosin}
{Sosin} C.,  1997, \mn@doi [\aj] {10.1086/118581}, \href
  {https://ui.adsabs.harvard.edu/abs/1997AJ....114.1517S} {114, 1517}

\bibitem[\protect\citeauthoryear{{Stetson}}{{Stetson}}{1987}]{stetson}
{Stetson} P.~B.,  1987, \mn@doi [\pasp] {10.1086/131977}, \href
  {https://ui.adsabs.harvard.edu/abs/1987PASP...99..191S} {99, 191}

\bibitem[\protect\citeauthoryear{Subramaniam et~al.,}{Subramaniam
  et~al.}{2017}]{Subramaniam2017}
Subramaniam A.,  et~al., 2017, \mn@doi [ApJ] {10.3847/1538-3881/aa94c3}, 154,
  233

\bibitem[\protect\citeauthoryear{{Thomson}, {Dieball}, {Knigge}, {Long}  \&
  {Zurek}}{{Thomson} et~al.}{2010}]{m80}
{Thomson} G.~S.,  {Dieball} A.,  {Knigge} C.,  {Long} K.~S.,   {Zurek} D.~R.,
  2010, \mn@doi [\mnras] {10.1111/j.1365-2966.2010.16729.x}, \href
  {https://ui.adsabs.harvard.edu/abs/2010MNRAS.406.1084T} {406, 1084}

\bibitem[\protect\citeauthoryear{{Thomson} et~al.,}{{Thomson}
  et~al.}{2012}]{ngc6752}
{Thomson} G.~S.,  et~al., 2012, \mn@doi [\mnras]
  {10.1111/j.1365-2966.2012.21104.x}, \href
  {https://ui.adsabs.harvard.edu/abs/2012MNRAS.423.2901T} {423, 2901}

\bibitem[\protect\citeauthoryear{Tody}{Tody}{1986}]{iraf1}
Tody D.,  1986, \mn@doi [SPIE] {10.1117/12.968154}, 0627, 733

\bibitem[\protect\citeauthoryear{{Tody}}{{Tody}}{1993}]{iraf2}
{Tody} D.,  1993, ASPCS, \href
  {https://ui.adsabs.harvard.edu/abs/1993ASPC...52..173T} {52, 173}

\bibitem[\protect\citeauthoryear{Tout \& Hall}{Tout \& Hall}{1991}]{tout1991}
Tout C.~A.,  Hall D.~S.,  1991, \mn@doi [MNRAS] {10.1093/mnras/253.1.9}, 253, 9

\bibitem[\protect\citeauthoryear{Tout, Pols, Eggleton  \& Han}{Tout
  et~al.}{1996}]{tout}
Tout C.~A.,  Pols O.~R.,  Eggleton P.~P.,   Han Z.,  1996, \mn@doi [MNRAS]
  {10.1093/mnras/281.1.257}, 281, 257

\bibitem[\protect\citeauthoryear{{Wang} et~al.,}{{Wang}
  et~al.}{2016}]{wang2016}
{Wang} L.,  et~al., 2016, \mn@doi [\mnras] {10.1093/mnras/stw274}, \href
  {https://ui.adsabs.harvard.edu/abs/2016MNRAS.458.1450W} {458, 1450}

\bibitem[\protect\citeauthoryear{Wood}{Wood}{1995}]{wood}
Wood M.,  1995, Theoretical white dwarf luminosity functions: DA models,
  \textit{White Dwarfs}.
Springer Berlin Heidelberg, p.~41

\bibitem[\protect\citeauthoryear{Xin, Ferraro, Lu, Deng, Lanzoni, Dalessandro
  \& Beccari}{Xin et~al.}{2015}]{xin}
Xin Y.,  Ferraro F.~R.,  Lu P.,  Deng L.,  Lanzoni B.,  Dalessandro E.,
  Beccari G.,  2015, \mn@doi [ApJ] {10.1088/0004-637x/801/1/67}, 801, 67

\bibitem[\protect\citeauthoryear{{Zurek}, {Knigge}, {Maccarone}, {Dieball}  \&
  {Long}}{{Zurek} et~al.}{2009}]{zurek2009}
{Zurek} D.~R.,  {Knigge} C.,  {Maccarone} T.~J.,  {Dieball} A.,   {Long} K.~S.,
   2009, \mn@doi [\apj] {10.1088/0004-637X/699/2/1113}, \href
  {https://ui.adsabs.harvard.edu/abs/2009ApJ...699.1113Z} {699, 1113}

\bibitem[\protect\citeauthoryear{Zurek, Knigge, Maccarone, Pooley, Dieball,
  Long, Shara  \& Sarajedini}{Zurek et~al.}{2016}]{zurek}
Zurek D.~R.,  Knigge C.,  Maccarone T.~J.,  Pooley D.,  Dieball A.,  Long
  K.~S.,  Shara M.,   Sarajedini A.,  2016, \mn@doi [MNRAS]
  {10.1093/mnras/stw1190}, 460, 3660

\makeatother
\end{thebibliography}

\label{lastpage}

\end{document}